\documentclass[twocolumn]{aastex62}

\newcommand{\Edit}{}

\newcommand{\moon}{\leftmoon\hspace{-0.19em}}
\newcommand{\eri}{$\epsilon$ Eridani}
\newcommand{\ceti}{$\tau$ Ceti}

\usepackage{wasysym}				
\usepackage{multirow}				
\usepackage{subfigure}				
\usepackage{verbatim}				

\received{2018 October 8}
\revised{2019 March 24}
\accepted{2019 April 15}
\submitjournal{ApJ}

\shorttitle{Gamma rays from Epsilon Eridani}
\shortauthors{Riley et al.}

\begin{document}

\title{Possible detection of gamma rays from Epsilon Eridani}

\correspondingauthor{A.H.~Riley}
\email{alexriley@tamu.edu}

\author[0000-0001-5805-5766]{Alexander H.~Riley}
\affiliation{Department of Physics and Astronomy, Mitchell Institute for Fundamental Physics and Astronomy, Texas A\&M University, College Station, TX 77843, USA}

\author[0000-0001-5672-6079]{Louis E.~Strigari}
\affiliation{Department of Physics and Astronomy, Mitchell Institute for Fundamental Physics and Astronomy, Texas A\&M University, College Station, TX 77843, USA}

\author[0000-0002-2621-4440]{Troy A. Porter}
\affiliation{Kavli Institute for Particle Astrophysics and Cosmology, Physics Department, Stanford University, Stanford, CA 94305 USA}

\author{Roger D. Blandford}
\affiliation{Kavli Institute for Particle Astrophysics and Cosmology, Physics Department, Stanford University, Stanford, CA 94305 USA}

\author{Simona Murgia}
\affiliation{Department of Physics and Astronomy, University of California, Irvine, CA 92697, USA}

\author[0000-0002-0893-4073]{Matthew Kerr}
\affiliation{Space Science Division, Naval Research Laboratory, Washington, DC 20375-5352, USA}

\author[0000-0003-1458-7036]{Gu{\dh}laugur J{\'o}hannesson}
\affiliation{Science Institute, University of Iceland, IS-107 Reykjavik, Iceland and Nordita,
KTH Royal Institute of Technology and Stockholm University Roslagstullsbacken 23, SE-106 91 Stockholm, Sweden}

\begin{abstract}
We use the {\it Fermi}-LAT gamma-ray observatory to search for gamma-ray emission from four nearby, debris disk-hosting main sequence stars: $\tau$ Ceti, $\epsilon$ Eridani, Fomalhaut, and Vega. For three stars ($\tau$ Ceti, Fomalhaut, and Vega), we establish upper limits that are consistent with theoretical expectations. For $\epsilon$ Eridani, we find a possible spatially coincident source with a soft energy spectrum of $dN/dE \sim E^{-3.6}$. However, at this stage we are unable to rule out that this emission is due to a more extended feature in the diffuse background. In the interpretation that the emission is due to $\epsilon$ Eridani, the $> 100$ MeV gamma-ray luminosity is $\sim 10^{27}$ erg/s $\simeq 3\times 10^{-7}$ L$_\odot$, which is $\sim 10^{10}$ times the gamma-ray luminosity from the disk of the quiet Sun. We find $\lesssim 2 \sigma$ evidence of source variability over a $\sim 7$ year timescale. In the interpretation that the gamma-ray emission from $\epsilon$ Eridani itself, we consider two possible models: 1) cosmic-ray collisions with solid bodies in the debris disk which extends out $\sim$60 AU from the host star, and 2) emission from the stellar activity. For the former model, assuming a total disk mass consistent with infrared measurements, we find that the size distribution of bodies is steeper than expected for a collisional cascade state. If confirmed as being associated with $\epsilon$ Eridani, this would be the first indication of gamma-ray emission from the vicinity of a main sequence star other than the Sun.
\end{abstract}


\section{Introduction} 
\label{sec:introduction}

\par The Sun and the Moon are well-studied gamma-ray emitters. The steady-state gamma-ray emission from the Sun is due to (1) the interactions of cosmic ray (CR) particles within and near the solar disk \citep[mostly involving hadronic processes;][]{1991ApJ...382..652S} and (2) through the rest of the heliosphere via Compton scattering of the solar photons by CR electrons \citep{2006ApJ...652L..65M,2007Ap&SS.309..359O}. Both components of the spectrum were first identified through analysis of EGRET data \citep{2008A&A...480..847O} and have been subsequently studied in greater detail using the {\it Fermi} Large Area Telescope~\citep[Fermi-LAT,][]{2011ApJ...734..116A,Ng:2015gya,2018PhRvL.121m1103L}. The spectral energy distribution from the solar disk is consistent with a single power law $dN/dE \propto E^{-\Gamma}$, with $\Gamma = 2.11 \pm 0.73$~\citep{2011ApJ...734..116A}. The flux $>$100\,MeV from inverse Compton (IC) emission within a $\sim20^\circ$ region around the Sun ($\sim 4.6 \times 10^{-7}$ cm$^{-2}$ s$^{-1}$) is similar to that from the disk ($\sim 6.8 \times 10^{-7}$ cm$^{-2}$ s$^{-1}$).

\par The high-energy gamma-ray emission from the Moon is due to CR interactions in the lunar regolith \citep{1984JGR....8910685M,2007ApJ...670.1467M}. The Moon was first detected in high-energy gamma rays by EGRET on CGRO~\citep{1997JGR...10214735T} and has been studied with greater sensitivity with the Fermi-LAT \citep{2012ApJ...758..140A,2016PhRvD..93h2001A}. The gamma-ray flux $>$100\,MeV from the Moon detected by the Fermi-LAT during a period of solar minimum conditions was $F_{\moon}(>$100$\,{\rm MeV}) = 1.04\pm0.01\pm0.1 \times 10^{-6}$ cm$^{-2}$~s$^{-1}$, a factor $\sim2-3$ larger than that obtained by EGRET in the 1990s during a period of higher solar activity. For energies $\gtrsim 200$~MeV the spectral energy distribution has index $\Gamma \sim 3.5$, softer than that of the Solar disk. This is due to the more rapid CR shower development in the denser materials of the lunar surface layers. The calculations made by~\citet{2007ApJ...670.1467M} provide a good description of the observed lunar flux for energies $\lesssim 200$~MeV for the central part of the disk and the rim. However, the observed spectrum of the rim is slightly harder than predicted. The difference between the predication and observation may be due to the roughness of the lunar surface which was not considered by \citet{2007ApJ...670.1467M}.

\par Aside from the Sun, the only other types of stars that are known gamma-ray emitters are early-type OB and Wolf-Rayet stars. If stars of this type are part of a binary or multi-star system, strong interacting winds generate non-thermal radiation, including gamma rays from IC emission or pion decay. Indeed, a source associated with the colliding-wind binary star Eta Carinae is evident in the Fermi-LAT data~\citep{2010ApJ...723..649A}. More generally, high-luminosity OB stars may be gamma-ray sources due to IC scattering of stellar photons by CR electrons~\citep{2007Ap&SS.309..359O}--- the same mechanism causing the solar heliospheric IC emission \citep[e.g.,][]{2006ApJ...652L..65M,2007Ap&SS.309..359O,2008A&A...480..847O} --- though sources of this class are yet to be detected with Fermi-LAT data.

\par In this paper, we consider the possibility that gamma-ray emission arises from the debris disks that surround nearby main sequence stars. These debris disks are comprised of both solid bodies with sizes $\sim 1$m-$100$ km and dust grains that orbit their host star~\citep{2018ARA&A..56..541H}. The dust is not left over from the star formation process, but rather is due to planetesimals that are stirred within the disk and collide to break up and produce the dust. The dust is identified as an excess of long wavelength sub-millimeter infrared (IR) emission in the spectral energy distribution of the star, relative to a black-body spectrum at the same stellar temperature. The wavelength of the excess corresponds to the approximate size of the dust grains in the debris disk. 

\par A debris disk thus is indicative of a population of bodies analogous to the asteroid and Kuiper belts in the Solar System. Gamma rays should be created through cosmic ray interactions with the planetesimals in a debris disk via similar processes to those that create the observed gamma rays from the Moon. It has been speculated that gamma rays may be produced similarly in the asteroid belt and the Oort cloud~\citep{2008ApJ...681.1708M,2009ApJ...692L..54M}, though this emission has not been identified in gamma-ray data. 

\par In this paper we perform the first search for, and analysis of, gamma rays from nearby main sequence stars. We generate baseline gamma-ray flux predictions from nearby debris disks by extrapolating from their measured IR fluxes, then use these predictions to generate a candidate target list of four nearby stars. We model the stars as point sources of gamma rays to test for significant emission of gamma rays and estimate the flux. We discuss systematics that may affect the identification of these sources, in particular focusing on contamination from low energy Galactic diffuse and isotropic gamma-ray emission. 

\par This paper is organized as follows. In Section~\ref{sec:debrisdisks} we review the basic properties of debris disks and estimate the gamma-ray flux that we expect from several nearby stars. In Section~\ref{sec:data} we describe the analysis of our sample of stars with the Fermi-LAT data. In Section~\ref{sec:results} we present the main results of our analysis, and in Section~\ref{sec:discussion} we provide a possible interpretation of the results. In Section~\ref{sec:conclusions} we present our conclusions. 

\section{Debris disks} 
\label{sec:debrisdisks}

\par In this section we obtain gamma-ray flux predictions for nearby stars with known debris disks. We begin with a brief review of debris disks, then discuss how this information may be used for flux predictions.  

\subsection{IR emission from debris disks}

\begin{table*}
\begin{center}
\caption{Properties of the stars and their debris disks: distance, spectral type, stellar effective temperature, fractional IR luminosity ($f = L_{IR}/L_\star$), mean radius of the disk ($r$), and modeled total mass of the disk from IR observations ($M$).  Where two values are listed, we show the quantities for the inner (outer) detected belt. 
\label{tab:properties}}
\begin{tabular}{cccccccccrrrrrrr}
\tableline\tableline
Star & Distance & Spectral & Temp. &  $f$ & $r$ & $M$ \\
& [pc] & Type & [K] & & [AU] & [M$_\oplus$] \\
\tableline
\multirow{2}{*}{\eri{}} & \multirow{2}{*}{3.2} & \multirow{2}{*}{K2V} & \multirow{2}{*}{5084} & \multirow{2}{*}{$3 \times 10^{-5} (4 \times 10^{-5})$~\tablenotemark{a,b}} & 14 (61)~\tablenotemark{a}& 0.4~\tablenotemark{a} \\
&  &  & & & 3 (63)~\tablenotemark{b} & 11~\tablenotemark{b} \\
\ceti{} & 3.65 & G8V & 5344 & $5 \times 10^{-5}$~\tablenotemark{c}  & 30~\tablenotemark{c,d} & 1.2~\tablenotemark{d} \\
Fomalhaut & 7.7 & A4V & 8590 & $2 \times 10^{-5} (8 \times 10^{-5})$~\tablenotemark{e} & 10 (140)~\tablenotemark{e,f} & 30~\tablenotemark{g} \\
Vega & 7.8 & A0Va & 9602 & $7 \times 10^{-6} (2 \times 10^{-5})$~\tablenotemark{e}  & 14 (140)~\tablenotemark{e}   & 10~\tablenotemark{h} \\
\tableline
\end{tabular}
\tablecomments{(a)~\citet{2014ApJ...791L..11G}; (b)~\citet{2009ApJ...690.1522B};
(c)~\citet{2014MNRAS.444.2665L}; (d)~\citet{2004MNRAS.351L..54G}; (e)~\citet{2013ApJ...763..118S}; (f)~\citet{2004ApJS..154..458S}; (g)~\citet{2002MNRAS.334..589W}; (h)~\citet{2010ApJ...708.1728M}}
\end{center}
\end{table*}

\par A debris disk can be described by its fractional luminosity, $f$, which is defined as the ratio of the $\sim$10$\mu$m$-1$mm IR luminosity re-radiated from dust to the bolometric optical luminosity of the star, $f \equiv L_{IR}/L_\star$. Typical measured values for the fractional luminosity are $f \sim 10^{-6} - 10^{-3}$~\citep{2008ARA&A..46..339W,2010RAA....10..383K,2014prpl.conf..521M}. To derive a relation between the measured parameter $f$ and the population of larger mass bodies which collide to form the observed dust, we must have a simple model for the debris disk. This model requires a description of the emission properties of the dust and a parameterization of the dust and planetesimal size distribution.

\par The intensity and spectrum of the host starlight that is absorbed and re-emitted by the dust in the disk depends on the size distribution of the dust. Assuming the dust is in thermal equilibrium, there is a balance between the energy absorbed and that which is emitted by the dust. To estimate the emission spectrum we assume that the dust acts as a perfect black body. This assumption is a theoretical simplification, as some observations indicate that the dust is at a higher temperature than a black body given its distance from the host star (see discussion in e.g.~\citet{2014prpl.conf..521M}). For the black-body assumption, the fractional luminosity of a disk in terms of the total cross sectional area of particles $\sigma_{tot}$ is 
\begin{equation}
f = \frac{\sigma_{tot}}{4\pi r^2}, 
\end{equation}
where $r$ is the distance from the center of the disk to the host star. Typical disk radii are $r \sim 10-100$ AU.

\par Modeling the size distribution as $dN/dD$, the total mass in bodies with diameters $D_1$ to $D_2$ is
\begin{equation} 
M (D_1,D_2) = \int_{D_1}^{D_2} \frac{\pi}{6} \rho D^3 \frac{dN}{dD} dD, 
\end{equation} 
where $\rho$ is the average internal density, and the total cross sectional area for bodies with diameters in this same range is 
\begin{equation} 
\sigma (D_1,D_2) = \int_{D_1}^{D_2} \frac{\pi}{4} D^2 \frac{dN}{dD} dD. 
\end{equation} 

\par The size distribution of bodies in the disk is typically assumed to follow a power law
\begin{equation}
\frac{dN}{dD} \propto D^{2 - 3k},
\label{eq:size}
\end{equation} 
where $k$ is a constant. This assumed functional form extends from the smallest-size dust grain ($\sim 1\mu$m, corresponding to the blowout radius at which dust is removed due to radiation pressure) to the largest-size planetesimal ($\sim$10s of km) that exists in the system. There are two important limiting cases that the distribution in Equation~\ref{eq:size} describes. For $k$ less than (greater than) 2, the mass is dominated by the largest planetesimals (smallest dust particles) in the distribution. For $k$ less than (greater than) 5/3, the cross sectional area is dominated by the largest planetesimals (smallest dust particles). 

\par The collisional cascade is a well-motivated theoretical model for the size distribution. It is defined as a steady state in which a given size bin loses particles to collisions at the same rate that it is replenished by the break up of even larger bodies~\citep{1969JGR....74.2531D}. For a collisional cascade, $k = 11/6$. In the collisional cascade model, the total mass of all bodies in a debris disk is given as~\citep{2008ARA&A..46..339W} 
\begin{equation}
M \simeq \frac{f}{0.37} r^2 D_{min}^{0.5} D_{max}^{0.5},
\label{eq:mtot}
\end{equation}
where the radius $r$ is in AU, $D_{min}$ is the minimum size body in $\mu$m, $D_{max}$ is the maximum size body in km, and $M$ is in units of Earth mass. Equation~\ref{eq:mtot} is typically used to estimate the total mass of bodies in debris disks; below we use this to compare to the gamma-ray observations. 

\par It is interesting to compare the predictions of the collisional cascade model to a simpler model for the total mass of a debris disk in which all of the bodies in the disk have the same diameter, $D_0$, so that the diameter distribution is $dN/dD \propto \delta (D - D_0)$. The cross sectional area of a body is $\sigma = \pi D_0^2/4$ such that $f = ND_0^2/16r^2$, where $N$ is the total number of bodies in the disk. The mass of an individual body is then $m = \rho \pi D_0^3/6$ and the assumed constant density of the body is $\rho$. In this case, the total mass of the disk is simply $M = N m$. 

\subsection{Gamma-ray emission}
\label{subsec:gamma-emission}

\par The Galactic cosmic rays (GCRs) with energy $\ge 100$ MeV that collide with planetesimals and dust are thought to originate mostly within sources such as supernova remnants that accelerate protons, nuclei, and electrons to high energies~\cite{Blasi:2013rva}. Though the local spectrum on Earth of GCRs is well-measured, the GCR spectrum is expected to vary throughout the Galaxy due to, for example, inhomogeneous sources distributions or an inhomogeneous magnetic field. 

\par From the GCR spectrum, the gamma-ray spectrum from the interactions of GCRs with rocky planetesimals can be determined.~\citet{2007ApJ...670.1467M} and~\citet{2008ApJ...681.1708M} have utilized the local GCR spectrum as an input to simulate interactions with the surface of the Moon. These authors showed that the gamma-ray spectrum results from two components, one that originates from the rim of the body, and one that arises from the central disk. Below energies of about \Edit{a few hundred  MeV}, the gamma-ray spectra from the disk and the rim are similar, whereas above \Edit{these energies} the spectrum is dominated by emission from the rim~\citep{2007ApJ...670.1467M}. These predictions provide a good description of the observed lunar spectrum for gamma-ray energies below about 200 MeV~\citep{2012ApJ...758..140A}. At higher energies, the observed spectrum is harder than is predicted. The lunar spectrum is well-described by a broken power law with $\Gamma = 1.1$ below 200 MeV, and $\Gamma = 3.5$ above 200 MeV. Because the GCR interactions occur near the Moon surface, the gamma-ray spectrum from the Moon provides a good template for interactions with planetesimals of diameter larger than the penetration depth within the rim of about 1 meter.  

\par For a first simple estimate of the gamma-ray flux from a debris disk, we take the total mass of the disk to be made of spherical bodies with a diameter $D$. Since the flux from a spherical body is proportional to the diameter of the body, to obtain a flux from a body of diameter $D$ we scale the observed flux from the Moon. The flux received from a single body in a disk at distance $d$ from the Earth is then
\begin{equation}
	F_{body} = F_{\moon} \left( \frac{D}{D_{\moon}} \right) \left( \frac{d_{\moon}}{d} \right)^2,
\end{equation}
where $F_{\moon} = 1.0 \times10^{-6}$ cm$^{-2}$ s$^{-1}$ \citep{2012ApJ...758..140A}, $D_{\moon}$ denotes the diameter of the Moon, $d_{\moon}$ is the distance to the Moon, and $d$ is the distance to the star.  Assuming that the spherical bodies have a constant internal density $\rho = 2.7$ g cm$^{-3}$ \citep[as in][]{2008ARA&A..46..339W}, the mass of a single body is $m = {\pi D^3\rho}/{6}$ and the total number of bodies in a disk of mass $M$ is
\begin{equation}
	N = \frac{M}{m} = \frac{6M}{\pi D^3 \rho}.
\end{equation}
The total flux from a disk with $N$ bodies is then 
\begin{eqnarray}
F &=& F_{\moon} \left( \frac{D}{D_{\moon}} \right) \left( \frac{d_{\moon}}{d} \right)^2 \frac{6M}{\pi D^3 \rho} \nonumber \\
&\simeq& 2.5\times10^{-7} \text{ cm$^{-2}$ s$^{-1}$} \left( \frac{M}{M_\oplus}\right) \left( \frac{\text{m}}{D} \right)^2 \left( \frac{\text{pc}}{d}\right)^{2}.
\label{eqn:flux}
\end{eqnarray}

\par We can provide a more detailed estimate by asuming a power law for the size distribution. To calculate the flux from the planetesimal component, we assume that the flux from a single planetesimal in a debris disk can be obtained by scaling to the flux from the Moon~\citep{2008ApJ...681.1708M}, 

\begin{eqnarray}
F & = & F_{\moon} \left( \frac{d_{\moon}}{d} \right)^2
\int_{D_{min}^\gamma}^{D_{max}^\gamma} \frac{dN}{dD} \frac{D}{D_{\moon}} dD \label{eq:gamma-ray_flux} \nonumber \\
& = &  \frac{F_{\moon}}{R_{\moon}} \left( \frac{d_{\moon}}{d} \right)^2
\left(\frac{4\pi}{3} \rho \right)^{1-k} 3 K \int_{D_{min}^\gamma}^{D_{max}^\gamma} D^{3 - 3k} dD 
\end{eqnarray}
\noindent
where

\begin{eqnarray}
&& \int_{D_{min}^\gamma}^{D_{max}^\gamma} D^{3 - 3k} dD= \left\{
\begin{array}{ll}
\displaystyle \frac{(D_{max}^\gamma)^{4-3k} - (D_{min}^\gamma)^{4 -3k}}{4-3k}, & k \neq 4/3 \\
\displaystyle \ln(D_{max}^\gamma/D_{min}^\gamma ), & k = 4/3
\end{array} \right.\nonumber \\
&& K = M \left\{
\begin{array}{ll}
\displaystyle \left(\frac{4\pi}{3}\rho\right)^{k - 2}\frac{2-k}{(D_{max}^\gamma )^{6-3k} - ( D_{min}^\gamma)^{6-3k} }, 
& k \neq 2 \\
\displaystyle \frac{1}{3}\ln (D_{max}^\gamma /D_{min}^\gamma ) & k = 2
\end{array} \right. \nonumber 
\end{eqnarray}
\noindent
In the formula above, ($D_{min}^\gamma$, $D_{max}^\gamma$) are the minimum and maximum size of the solid bodies that contribute to the gamma-ray observations. 
For this analysis, we take $D_{min}^\gamma = 1$ m and $D_{max}^\gamma = 100$ km to be the minimum and maximum-size planetesimals in the distribution. 

\begin{figure*}
\centering
\subfigure[\eri{}]{\label{fig:eridani_cmap}
\includegraphics[width=.23\linewidth]{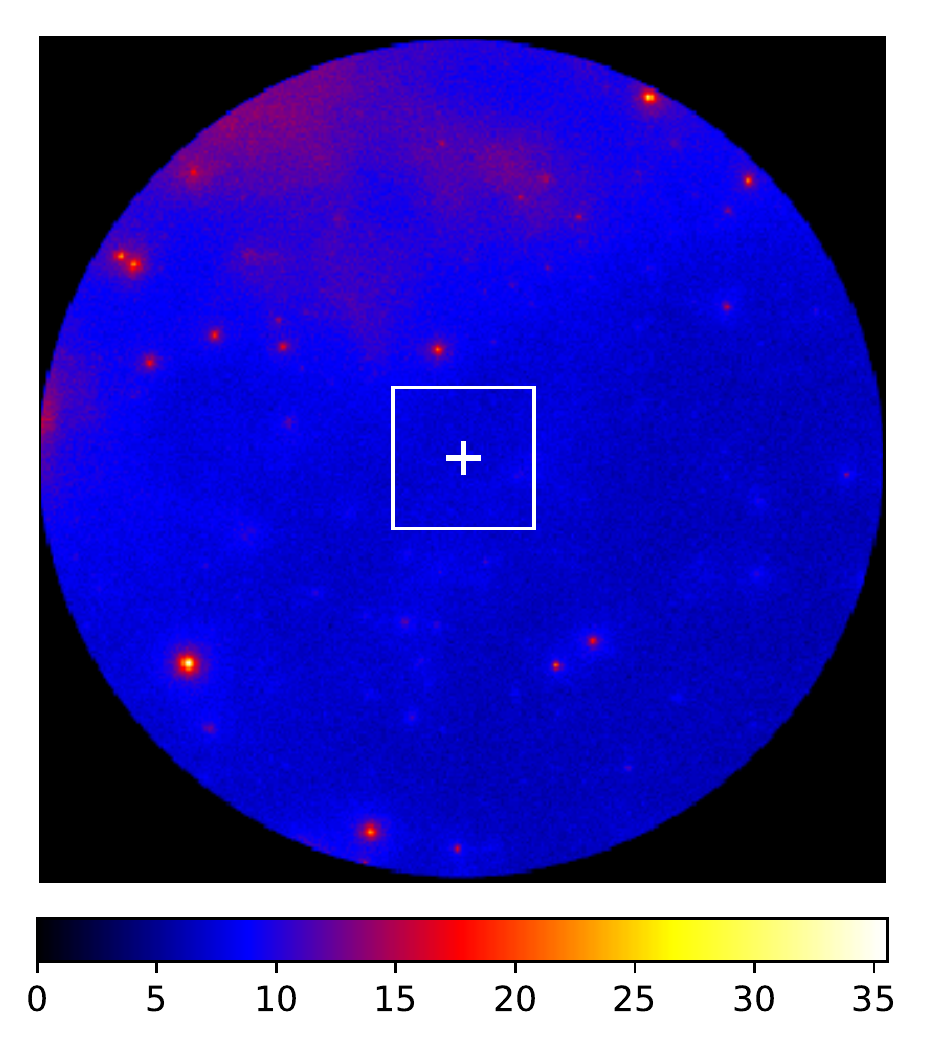}}
\subfigure[\ceti{}]{\label{fig:tau_cmap}
\includegraphics[width=.23\linewidth]{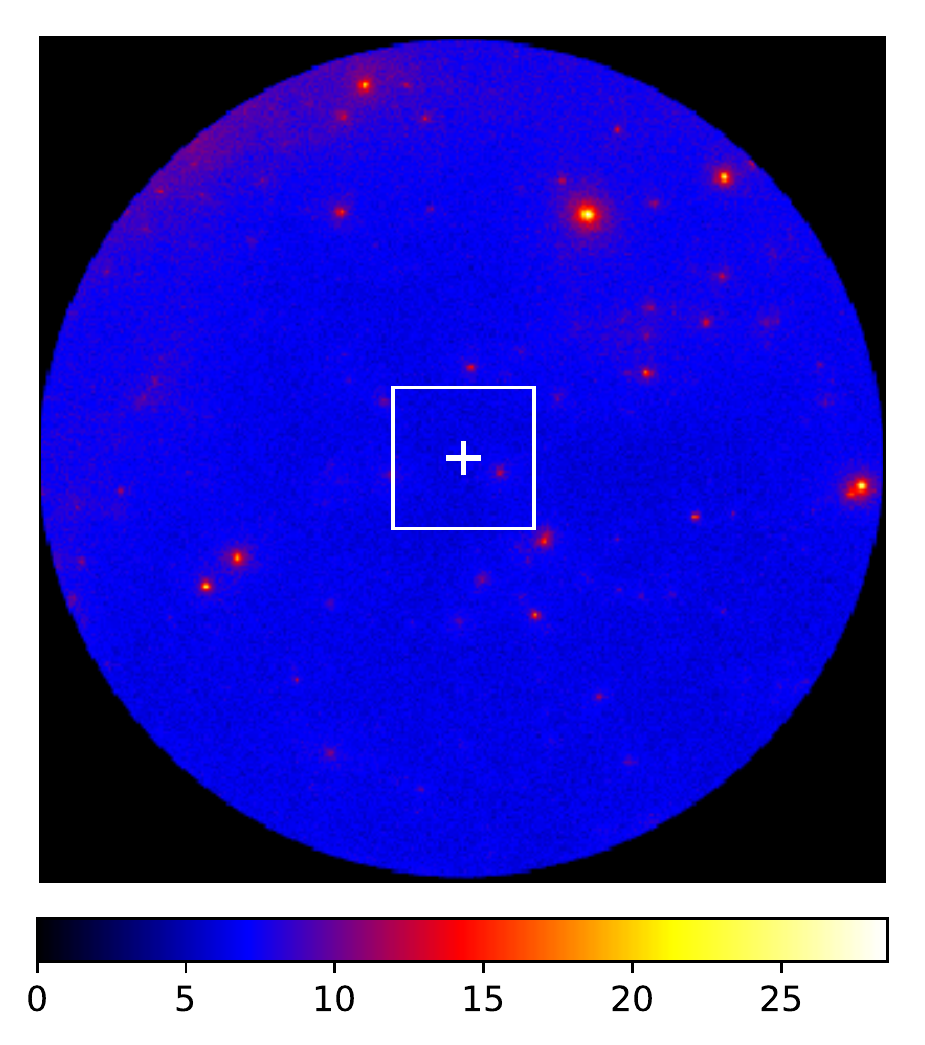}}
\subfigure[Fomalhaut]{\label{fig:fomalhaut_cmap}
\includegraphics[width=.23\linewidth]{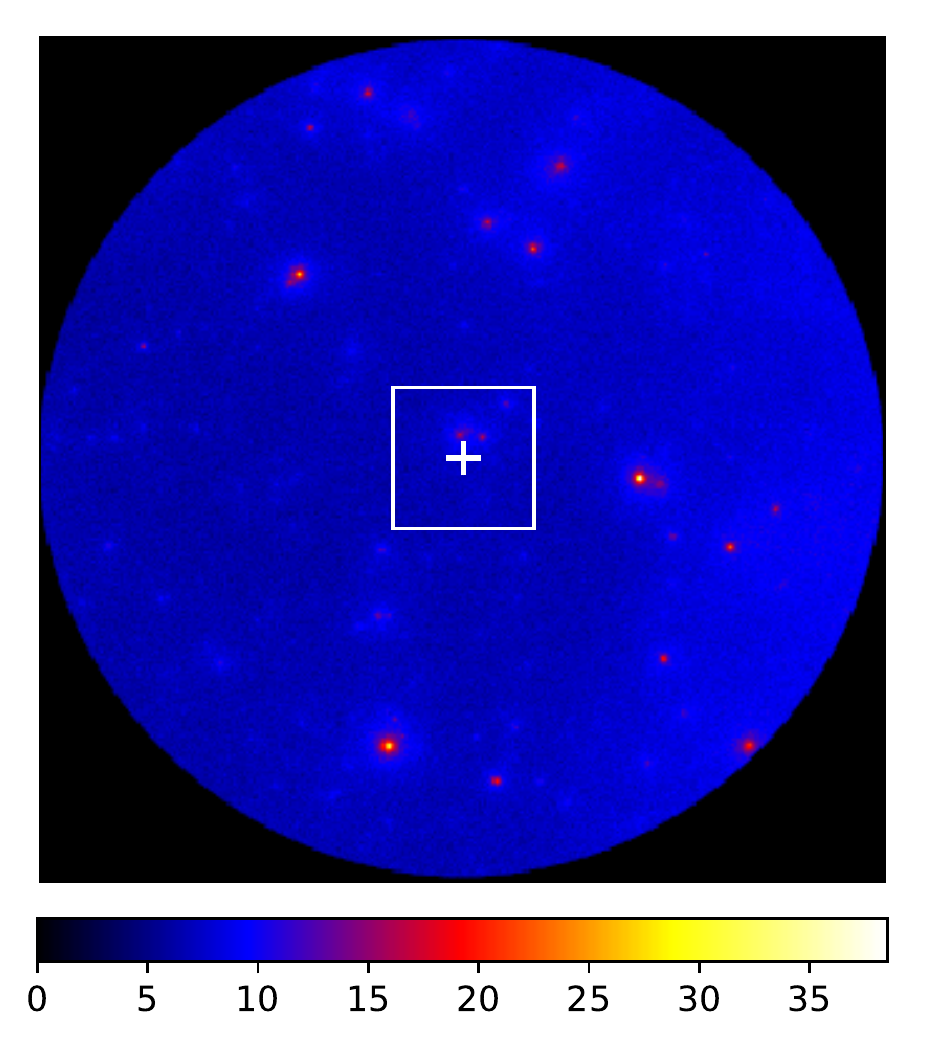}}
\subfigure[Vega]{\label{fig:vega_cmap}
\includegraphics[width=.23\linewidth]{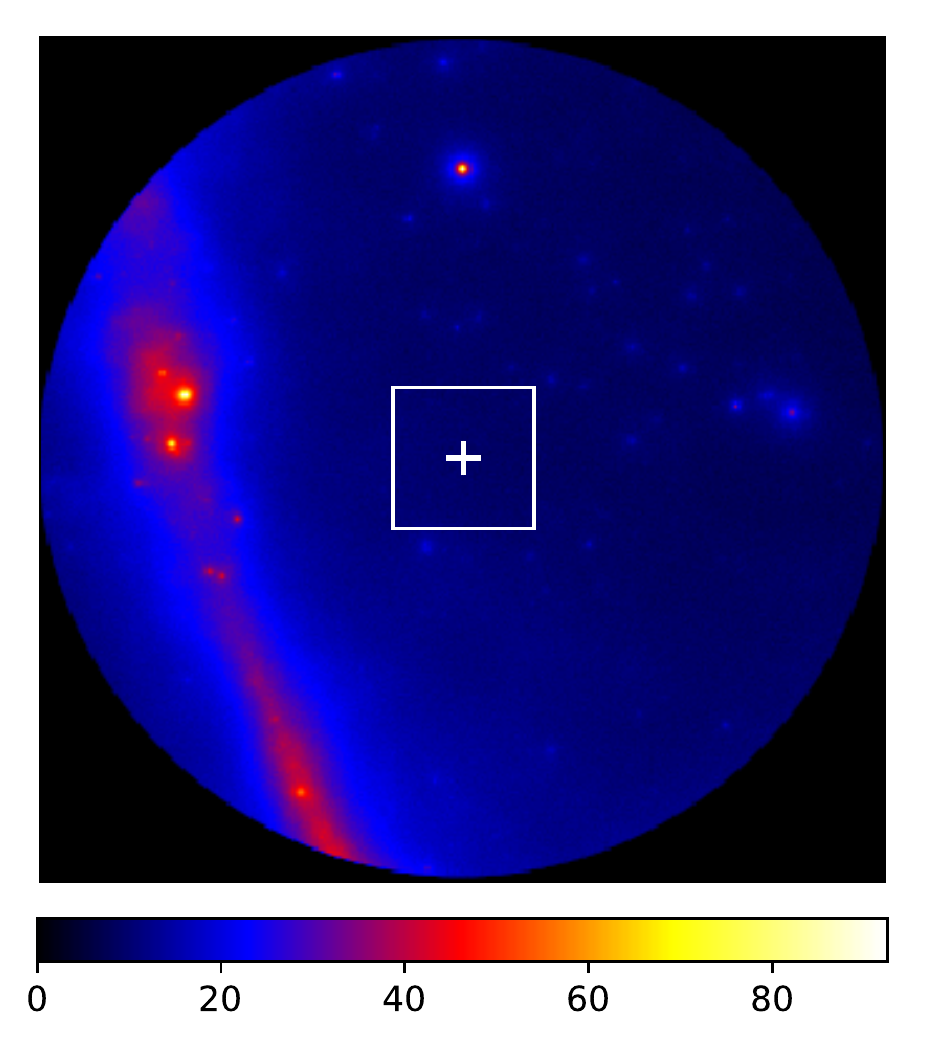}}
\subfigure{
\includegraphics[width=.23\linewidth]{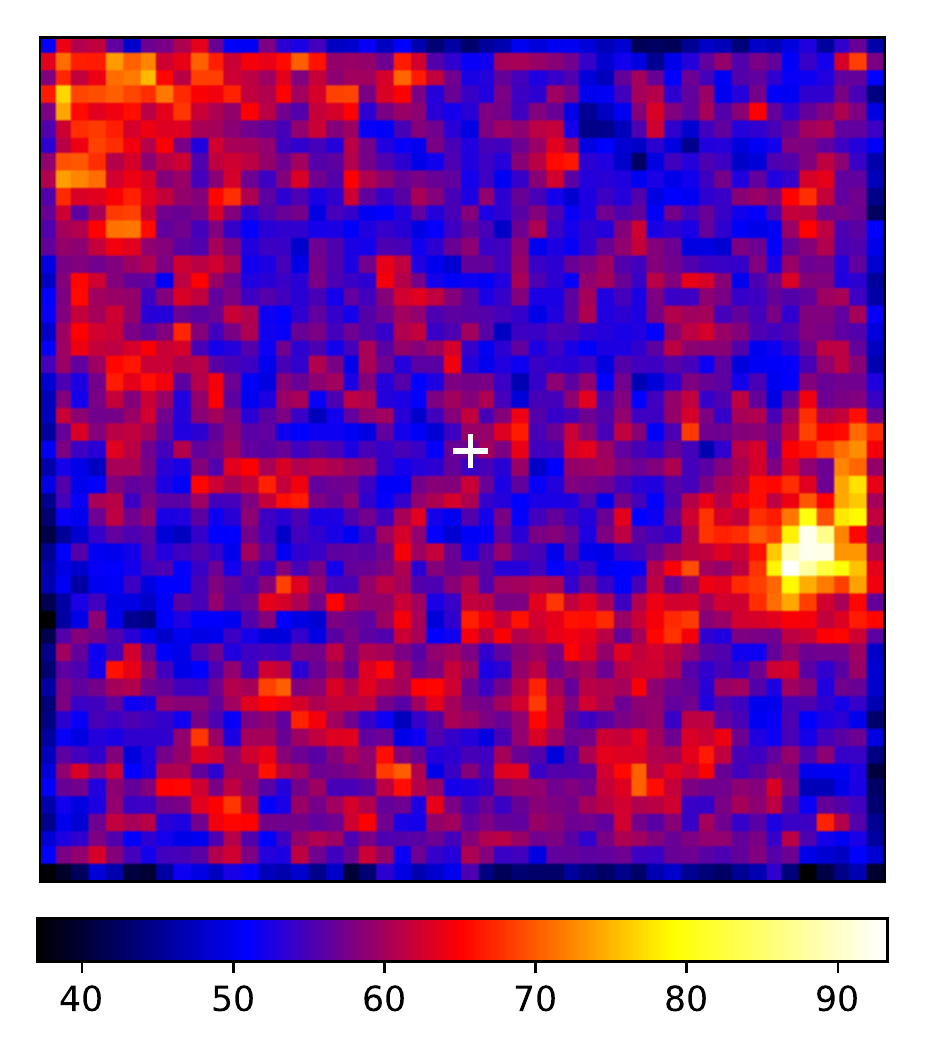}}
\subfigure{
\includegraphics[width=.23\linewidth]{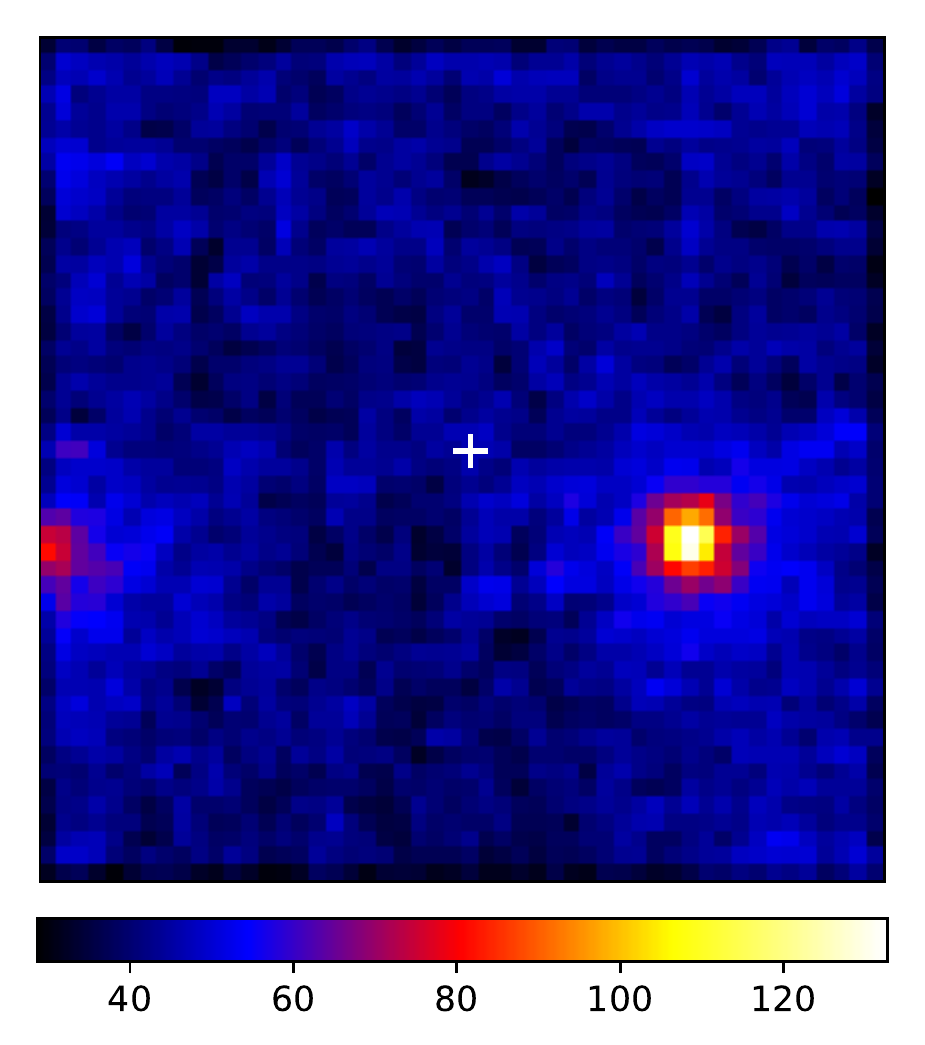}}
\subfigure{
\includegraphics[width=.23\linewidth]{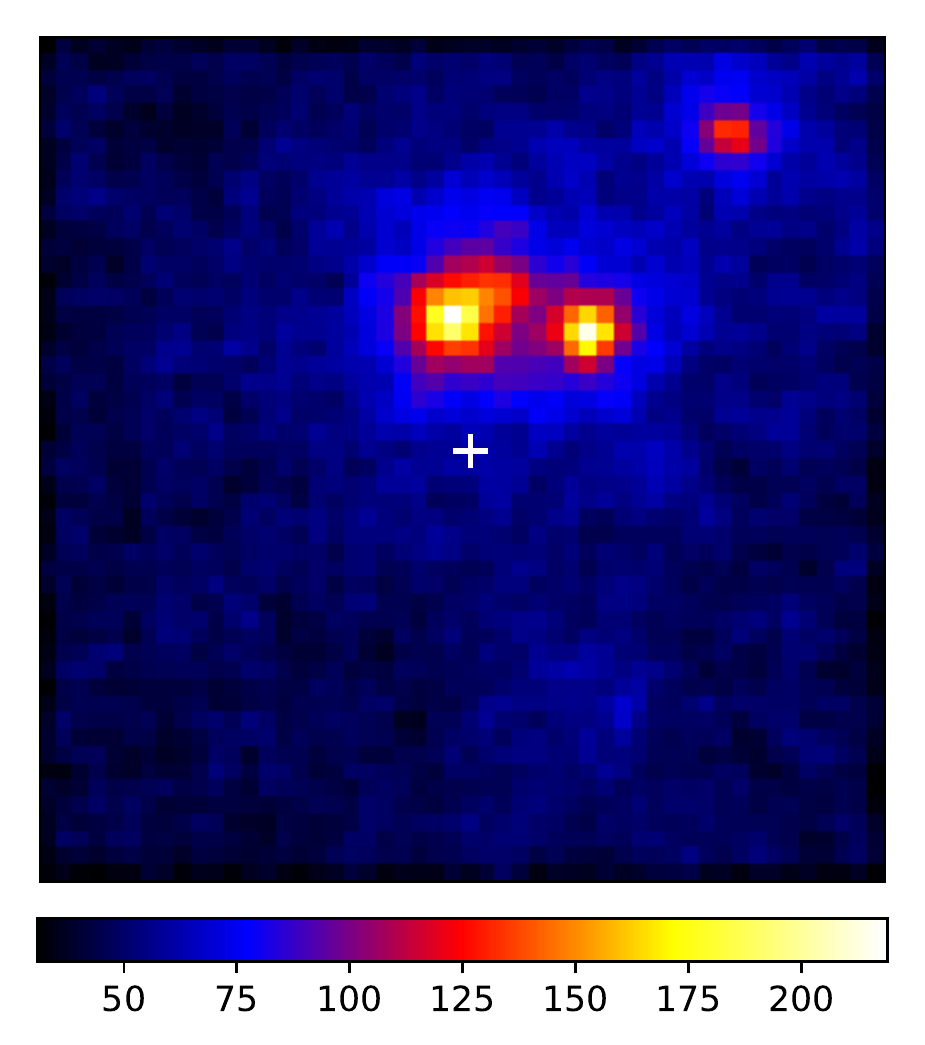}}
\subfigure{
\includegraphics[width=.23\linewidth]{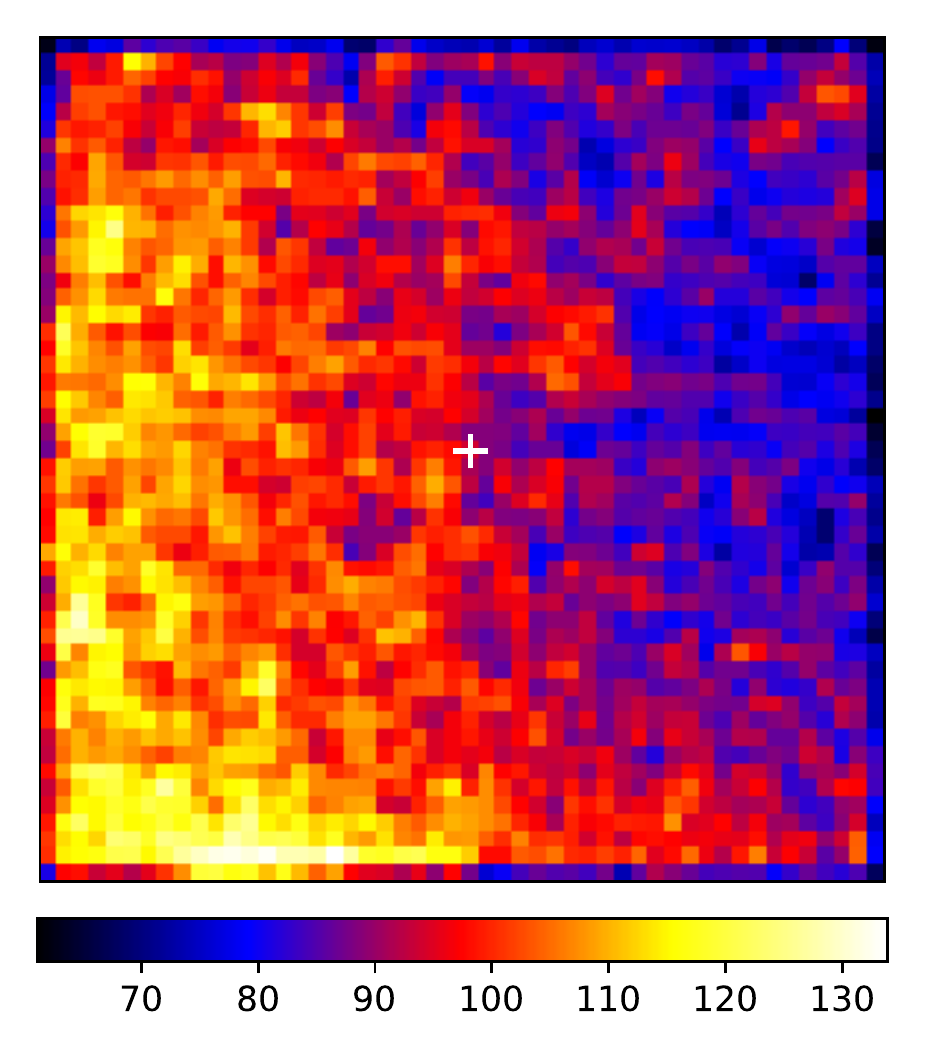}}
\caption{Counts maps for regions of interest with radius 30$^\circ$ around each of our four target stars (positions indicated by white crosses). \Edit{We also include maps zoomed on the star positions that are 5$^\circ$ on a side (positions of these zooms are indicated in the top row as white boxes)}. Photons with energies in the range 100\,MeV -- 300\,GeV are shown. The maps are based on 8.9 years of data. A square-root scaling is performed on each spatial bin \Edit{in the top row of images}, and the colorbar scales vary for each image. For Vega (panel d), the diffuse emission due to the Galactic plane is evident.}
\label{fig:cmaps}
\end{figure*}

\par We determine the normalization constant, $K$, by fixing the total mass of the combined population of dust and solid bodies in the disk between $D_{min}$ and $D_{max}$ as $M = M(D_{min}, D_{max})$. Here $D_{min}$ is the minimum size dust particle in the debris disk, and $D_{max}$ is the maximum size of the solid body. We take the minimum size dust to be equal to the size of the blowout radius at which dust is removed due to radiation pressure, and the maximum size solid body to be $D_{max} = D_{max}^\gamma$. Through this total mass, $M$, as well as through the spectral index $k$, the gamma-ray and IR observations are tied together. For a suitable set of parameters that describe $dN/dD$, we are able to calculate the IR and gamma-ray fluxes. 

\par Using the above formalism we can predict the gamma-ray fluxes from four nearby prominent debris disks: \eri{}, \ceti{}, Vega, and Fomalhaut. We choose these stars because they are the four nearest stars with known debris disks. The properties of these disks are listed in Table~\ref{tab:properties}, while Table~\ref{table:allstars} lists the predicted fluxes for the collisional cascade model and a model in which all of the bodies have a radius of 1 meter. 

\par \Edit{The predictions of~\citet{2007ApJ...670.1467M} are valid for high and low periods of solar activity, and an extrapolation based on predicted local interstellar spectra. The local interstellar spectra may be estimated from recent data.} To be specific, comparison of AMS-02 and Voyager data~\citep{2017PhRvD..96j3005T} shows the proton flux is suppressed below $\sim10$~GeV and the reduction at 1 AU  is already a factor $\sim 7$ at 1~GeV and very large at 100 MeV. \eri{} has a stellar wind with discharge about thirty times solar~\citep{2008MNRAS.385.1691N} and a correspondingly larger termination shock, and its debris disk extends out to $\sim100$~AU. \Edit{Although these details of the flux impacting the disk may be ultimately be used to make a more precise estimate of the flux from its disk, for simplicity we take the flux impacting the disk will be similar to that incident upon the moon and so we use the lunar spectrum for comparison purposes.}

\par In addition to the emission from the solid bodies in the disk, we note that gamma rays may be produced from interaction of GCRs with the dust in the debris disk. An estimation of the emissivity from this component involves detailed knowledge of the composition of the dust. Under the simple assumption that the dust is hydrogen nuclei, the $>$100\,MeV gamma-ray emissivity is related to the integrated flux over this same energy range as $F_{dust} = (M_{dust}/m_p) \times  q_\gamma /d^2$, where $d$ is the distance to the star, $M_{dust}$ is the total dust mass, $m_p$ is the proton mass, and $q_\gamma$ photons/(s H-atom) is the  gamma ray emissivity per hydrogen atom greater than 100 MeV. In this case the spectral energy distribution is expected to be harder than the spectral energy distribution from the solid bodies. The Fermi-LAT measured this gamma-ray emission from the local interstellar medium~\citep{2009ApJ...703.1249A, 2012ApJ...750....3A}. Since the dust in debris disks is composed of heaver nuclei, e.g. enhanced silicate features~\citep{2014prpl.conf..521M}, the gamma-ray emissivity must be corrected to account for the detailed nuclear composition.  

\section{Data analysis}
\label{sec:data}

\par To determine if the stars in our sample in Table~\ref{tab:properties} have gamma-ray counterparts, we perform a standard point-source binned likelihood analysis using the Fermi Science Tools' v11r5p3~\footnote{https://fermi.gsfc.nasa.gov/ssc/} \texttt{gtlike} function. We use \texttt{Pass 8 SOURCE}-class events with energies in the range 100 MeV to 300 GeV with over 8.9 years of data. We apply the recommended  \texttt{(DATA\char`_QUAL>0)\&\&(LAT\char`_CONFIG==1)} filter to ensure quality data and a zenith cut $z_{max} = 90^\circ$ to filter background gamma-ray contamination from the Earth's limb.  With these cuts the resulting counts map for each source is shown in Figure \ref{fig:cmaps}.

\par Using the data selections above, we bin the photons into 37 logarithmically spaced energy bins. We take a 0.2$^\circ$ angular pixelation and use the \texttt{NEWMINUIT} optimizer method within \texttt{gtlike}. We use an input source model that includes all sources in the 3FGL catalog within a Region Of Interest (ROI) of $30^\circ$ around each star, as well as within an additional 10 degrees to account for PSF overlap from sources outside of the ROI. The total number of 3FGL sources in each source model is: 253 sources for \eri{}, 228 for \ceti{}, 171 for Fomalhaut, and 300 for Vega.

\par Our source model also includes a point source at the center of the ROI representing the potential source star. We model the spectral energy distribution of the star two ways: (1) as a power law with a free spectral index and (2) as a fixed broken power law which is similar to the spectrum observed from the Moon. For the background sources and diffuse and isotropic backgrounds, we consider the following two approaches. In the first, we fix the spectral energy distributions and the flux normalizations for all background 3FGL point sources to their 3FGL values and leave free the normalizations of the isotropic and diffuse backgrounds (``BKGD" model in Table \ref{table:allstars}). Note that in the 3FGL analysis there is also spectral freedom in the diffuse emission model~\citep{2015ApJS..218...23A}, which we do not include here. In the second approach, in addition to the normalizations of the isotropic and diffuse backgrounds, we free the flux normalizations of the 3FGL point sources within $5^\circ$ of the star (``5DEG" model in Table \ref{table:allstars}). For the second approach, we keep fixed the spectral index of the sources within the ROI to their 3FGL values. 

\par As a result of each \texttt{gtlike} run centered around a potential source star, we obtain the Test Statistic ($TS$), which is indicative of how much support the model has for a source at the position of the star. The $TS$ value is approximately equal to the square of the detection significance for that source, with a nominal value of $TS = 25$ showing significant evidence of the detection of a point source. For the instances in which the $TS$ value is well below the detection threshold, we utilize the Upper Limits algorithm from the \texttt{pyLikelihood} implementation of the Fermi Science Tools to derive an upper limit on the gamma-ray flux from the star. 

\section{Results} 
\label{sec:results}

\par In this section we present the results obtained from the binned likelihood analysis procedure described above. We then undertake a more specific analysis in the region of \eri{}, which is our only source with a $TS$ value approaching significance.

\subsection{Binned likelihood results}

\par
\begin{figure}
\centering
\includegraphics[width=\linewidth]{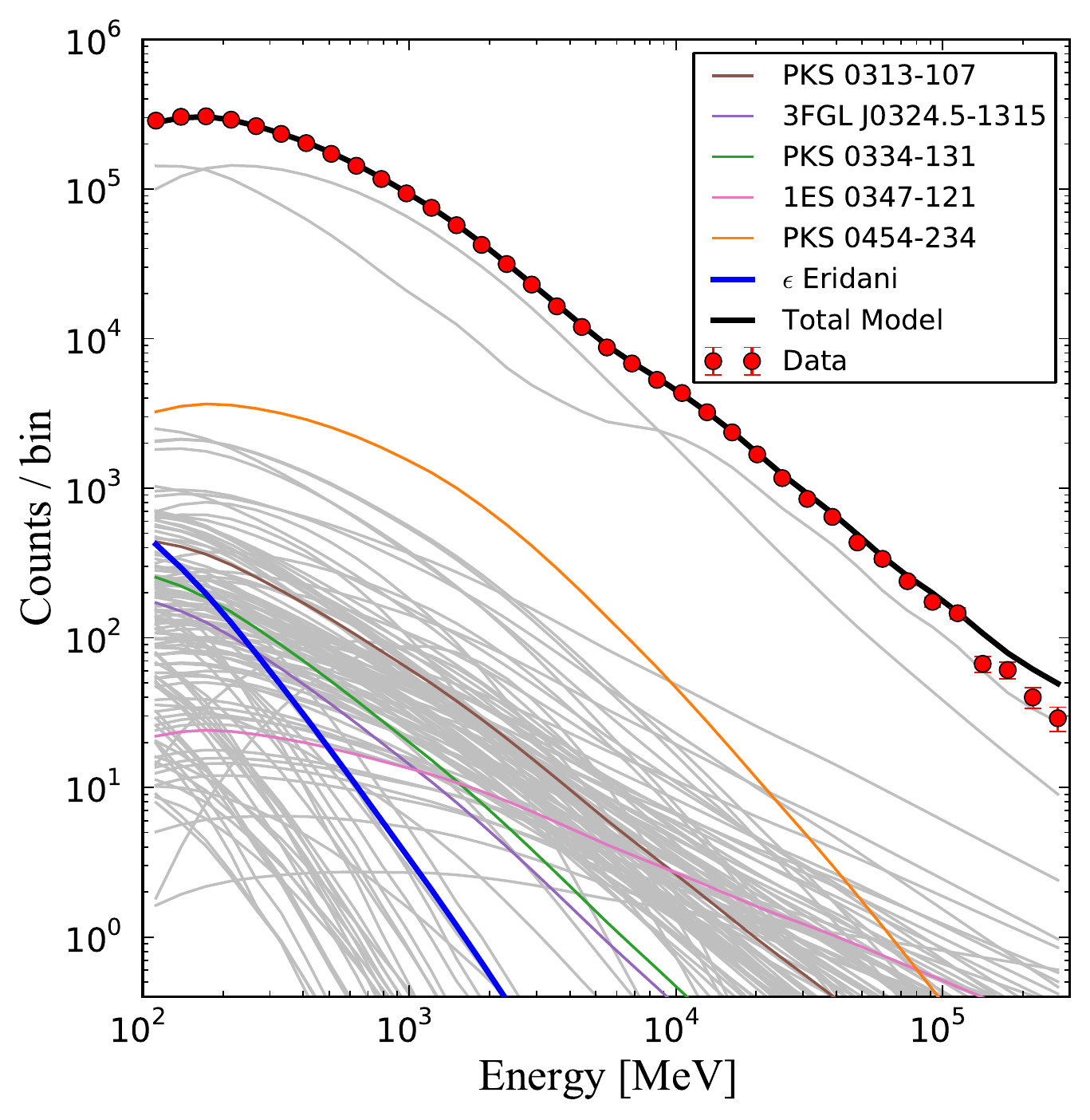}	
\caption{Counts distribution for the data and sources within the 30$^\circ$ ROI centered on \eri{}. Photons with energy in the range 100\,MeV -- 300\,GeV are shown. The top two grey lines are the diffuse Galactic and isotropic backgrounds, and the curves at the bottom left are the point sources within the ROI. The solid blue curve is the best fitting model for the \eri{} source. Note that error bars on the data are present, but are not discernible for energy bins $<10^5$ MeV.}
\label{fig:sed}
\end{figure}

 The results of our binned likelihood analyses are summarized in Table~\ref{table:allstars} for both the BKGD and 5DEG models. The comparison between the fluxes we obtain for 3FGL sources and those reported in the 3FGL catalog is shown in the Appendix. For \eri{}, we show the results for an assumed pure power law gamma-ray spectrum. We find that \texttt{gtlike} converges to a best fit value of $\Gamma \simeq 3.6 \pm 0.2$ \Edit{for energies $> 100$ MeV}, with a $TS = 31.7$ for the case in which sources within 5 degrees are varied, and $TS = 25.3$ for the case in which all 3FGL sources are fixed. Figure~\ref{fig:sed} shows the counts distribution for \eri{} as well as the diffuse emission and other sources within the ROI.

\par For the other three stars (\ceti{}, Fomalhaut, and Vega), for a pure power law spectrum we find that \texttt{gtlike} does not find associated gamma-ray emission with the respective locations. Therefore for these three sources, we quote flux upper limits using the Moon spectrum, which we model as a broken power law described in Section \ref{subsec:gamma-emission}.

\begin{table*}
\centering
	\begin{tabular}{cc||cccrc|}
    Source & Model & Flux & $F_{pred.}^{k=11/6}$ & $F_{pred.}^{R=1\text{m}}$ & $TS$ & Lumin. \\
    \hline \hline
    \multirow{2}{*}{\eri{}} & BKGD & $8.1\pm1.6$ & $8.3\times10^{-4}$ & \multirow{2}{*}{2.5 (68)} & 25.3 & $6.7\times10^{-7}\pm 1.4\times10^{-7}$ \\
    & 5DEG & $9.2\pm1.7$ & ($2.3\times10^{-2}$) & & 31.7 & $7.4\times10^{-7}\pm 1.4\times10^{-7}$ \\
    \hline
    \multirow{2}{*}{\ceti{}} & BKGD & $<$0.54 & \multirow{2}{*}{$1.9\times10^{-3}$} & \multirow{2}{*}{5.7} & $-1.8$ & $< 7.7\times10^{-8}$ \\
    & 5DEG & $<$0.82 & & & $-0.9$ & $< 1.1\times10^{-7}$ \\
    \hline
    \multirow{2}{*}{Fomalhaut} & BKGD & $<$0.39 & \multirow{2}{*}{$1.1\times10^{-2}$} & \multirow{2}{*}{32} & $-3.5$ & $< 2.4\times10^{-7}$ \\
    & 5DEG & $<$4.28 & & & 2.9 & $< 2.7\times10^{-6}$ \\
    \hline
    \multirow{2}{*}{Vega} & BKGD & $<$1.07 & \multirow{2}{*}{$3.5\times10^{-3}$} & \multirow{2}{*}{10} & $-0.6$ & $< 6.9\times10^{-7}$ \\
    & 5DEG & $<$5.54 & & & 2.8 & $< 3.6\times10^{-6}$ \\
    \hline
  	\end{tabular}
\caption{Results for the gamma-ray flux $>$100\,MeV (in units $10^{-9}$ cm$^{-2}$ s$^{-1}$) for each of our four target stars. For \eri{}, we show the results from the best-fitting pure power law spectrum and report a flux measurement. For the other three sources, we utilize the Moon spectrum, which is approximated as a broken power law, and report upper limits on the gamma-ray flux \Edit{and luminosity}. The \Edit{fourth and fifth columns} give our predicted gamma-ray flux ($>$100\,MeV) for the collisional cascade model, $k = 11/6$, and uniform spherical body model with $R=1$ m (note that \eri{} has two values due to two different mass estimates, see Table \ref{tab:properties}). \Edit{The final two columns give the $TS$ result and gamma-ray luminosity (in units L$_\odot$).} Under the ``Model" column, BKGD fixes all sources within the ROI, while varying only the diffuse Galactic and extragalactic backgrounds. The 5DEG label denotes that sources within 5 degrees are freed and fit for in addition to the diffuse Galactic and extragalactic backgrounds.} 
\label{table:allstars}
\end{table*}

\par Focusing on the potential gamma-ray source associated with \eri{}, we examine which energy range provides the most substantial contribution to the $TS$ value. We repeat the same binned likelihood analysis as above, only now increasing the lower energy threshold from 100 MeV, for each of the BKGD and 5DEG models. The results are summarized in Table~\ref{table:checks}, and for the sources within 5 degrees we also compare our 5DEG fluxes to those obtained in the 3FGL catalog in the Appendix. For an energy threshold $>$200\,MeV, the source has a $TS$ in the range $\sim 18-19$, with the precise value depending on how the background sources are modeled and whether $\Gamma$ is free or is allowed to vary. For energies $>$500\,MeV, the $TS$ for the source drops significantly, and for energies $>$1\,GeV the $TS$ is negligible.  

\begin{table}[t]
\centering
	\begin{tabular}{cc||ccc||}
    Data selection & Model & TS & Upper limit & $\Gamma_{fit}$ \\
    \hline \hline
    \multirow{2}{*}{$>$200 MeV} & BKGD & 18.4 & 1.60$\times10^{-8}$ & -3.385 \\
    & 5DEG & 19.5 & 1.69$\times10^{-8}$ & -3.427 \\
    \hline
    \multirow{2}{*}{$>$500 MeV} & BKGD & 6.81 & 1.18$\times10^{-8}$ & -2.968 \\
    & 5DEG & 6.86 & 1.20$\times10^{-8}$ & -2.974 \\
    \hline
    \multirow{2}{*}{$>$1 GeV} & BKGD & 2.31 & 3.84$\times10^{-9}$ & -2.479 \\
    & 5DEG & 2.32 & 6.20$\times10^{-7}$ & -2.480 \\
    \hline
    \multirow{2}{*}{\texttt{PSF0+PSF1}} & BKGD & 19.6 & 3.53$\times10^{-9}$ & -2.83 \\
    & 5DEG & 25.2 & 7.96$\times10^{-9}$ & -2.96 \\
    \hline
    \multirow{2}{*}{\texttt{PSF2+PSF3}} & BKGD & 7.9 & 7.05$\times10^{-9}$ & -3.0* \\
    & 5DEG & 10.0 & 8.81$\times10^{-9}$ & -3.0* \\
    \hline
  	\end{tabular}
\caption{Further analysis of the $TS$ for \eri{}. We make cuts on Fermi data depending on the assumed lower energy threshold and the PSF event type. For each cut, we vary the model for the background sources within the ROI. For both of the \texttt{PSF0+PSF1} and \texttt{PSF2+PSF3} tests, we take an energy threshold of $>$100\,MeV. * indicates spectral index fixed for convergence.
\label{table:checks}}
\end{table}

\par \Edit{We note that for the lower-energy cut $>$200\,MeV, as well as the fiducial $>$100\,MeV, the best-fitting spectral index for \eri{} is similar to that of the Moon over the same energy range ($\Gamma = 3.5$)}. This leads to the natural interpretation that the gamma-ray emission is due to solid bodies in the debris disk, and in section~\ref{sec:discussion} below we discuss this interpretation in more detail. Assuming a broken power law spectrum similar to that observed from the Moon \citep{2016ApJ...819...44A}, we have re-run the \texttt{gtlike} analysis described above. With this model we obtain $TS \sim 25.8$. For this assumed spectrum, the best-fitting flux and uncertainty is nearly identical to that from the pure power law fit.

\par In addition to checks on the energy threshold, we perform checks on the localization of the source. We utilize the new partition for photon events based on the quality of their point-spread function (PSF) which accompanies the Pass 8 release of Fermi data. We repeat the same binned likelihood analysis as described above, with a restriction of selecting events from the lower (\texttt{PSF0+PSF1}) and upper (\texttt{PSF2+PSF3}) two quartiles. 

\par The results are summarized in Table~\ref{table:checks}. For the \texttt{PSF0+PSF1} event cut, the range of TS is similar to the above results. However, for the high-quality \texttt{PSF2+PSF3} events, the TS drops to $\sim 8-10$. This may be indicative of a more extended feature in the diffuse emission in this region of the sky. 

\subsection{\eri{}}

\par The above likelihood analysis shows that it is possible that a gamma-ray source is consistent with the spatial location of \eri{}. For the remainder of this section, we provide further analysis on the characteristics of this emission.

\subsubsection{Simulated sources} 

\par Soft sources similar to that associated with \eri{}, with power law spectral index $\Gamma > 3$, are rare in 3FGL ($\lesssim 0.4\%$); both the mean and median spectral index for power law sources in 3FGL are $\Gamma \simeq 2.2$. In order to determine the prospects for reconstructing the flux and the $\Gamma$ values for soft sources, we perform simulations with \texttt{gtobssim} from the Fermi Science Tools. The goal of these simulations is to test the sensitivity of the analysis to recover faint sources using a known background, which is particularly important at low energies, and understand how well the $TS$ and $\Gamma$ values are reconstructed for soft sources. For our simulations we take a region of sky centered on \eri{}, though our analysis is general enough that it could be applied to any similar ROI. Along with the simulated \eri{} source, we simulate the diffuse Galactic and extragalactic backgrounds and 3FGL sources located within 10$^\circ$ of \eri{}. 

\par We simulate sources at the position of \eri{} for several input values of $\Gamma$, spanning harder sources with $\Gamma = 2.8$ to softer sources with $\Gamma = 3.7$. We also simulate a broken power law spectrum with the same spectral index values and energy break as the Moon's spectrum. For each simulation we fix the total flux of the source at 9.2$\times10^{-9}$ cm$^{-2}$ s$^{-1}$, which corresponds to the flux of the best-fit model above. For our assumed flux and $\Gamma > 2.8$ we find the reconstructed value is softer than the input value; for example for a true $\Gamma = 2.8$, we find the reconstructed value spectral index to be $\sim 7\%$ greater, and for $\Gamma = 3.8$, we find the reconstructed value spectral index to be $\sim 20\%$ greater. 

This indicates that for true sources with $\Gamma > 2.8$, the reconstructed spectral energy distribution may be biased relative to the true $\Gamma$. However in spite of this bias in the reconstruction of $\Gamma$, we find that the source is detected with a $TS > 25$ for the entire range of input spectra considered.

\subsubsection{PGWave}

\begin{figure}[t]
\centering
\includegraphics[width=\linewidth]{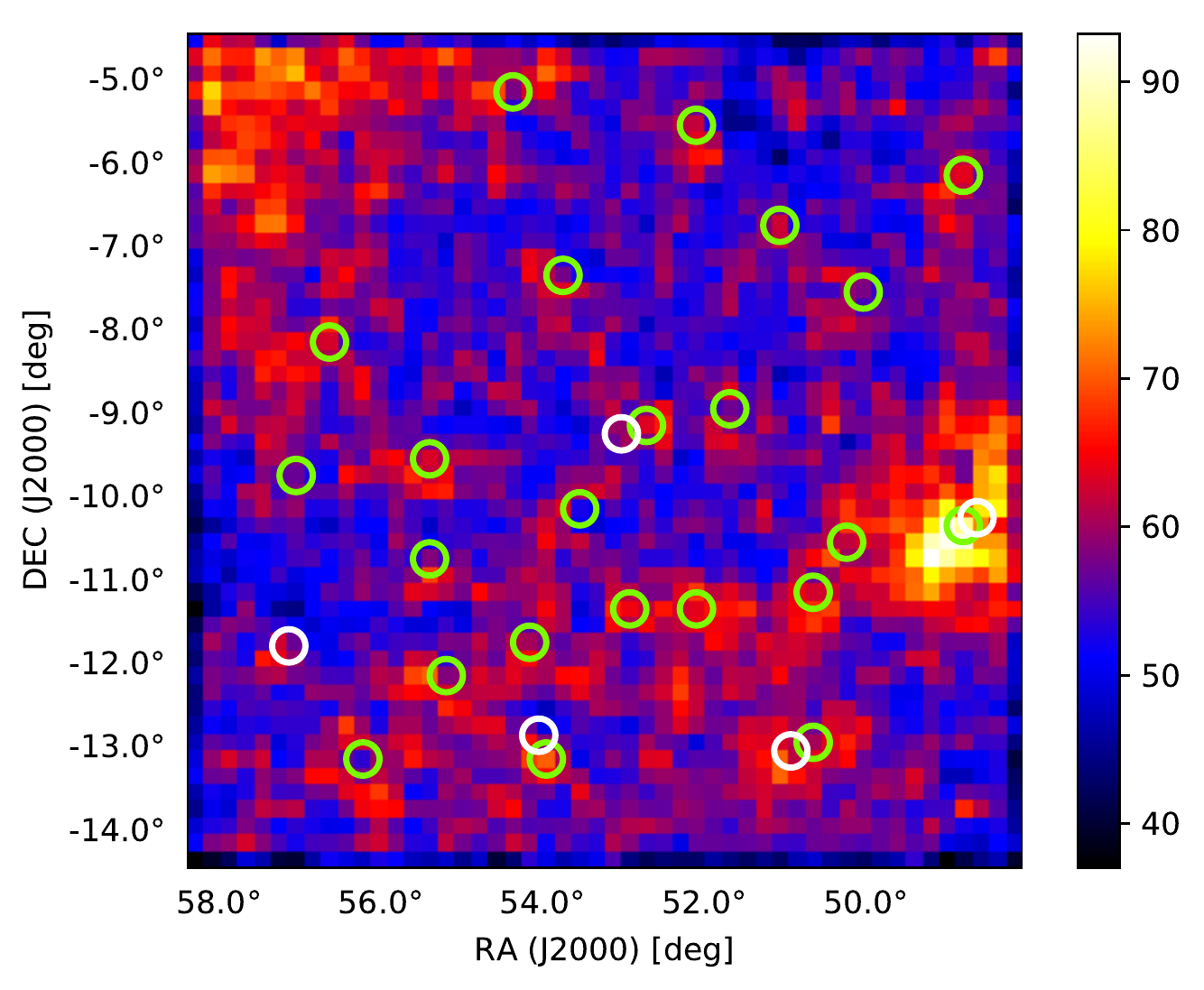}	
\caption{Counts map of {\it Fermi}-LAT data with combined seeds (green circles) from applying {\it PGWave} to the ROI around \eri{}. We are able to recover the location of the \eri{} source (central white circle), the locations of 3 out of 4 3FGL sources (other white circles), and identify some structures in the interstellar emission (see text for details).}
\label{fig:pgwave}
\end{figure}

\par As emphasized above, the best-fitting spectral energy distribution for \eri{} is softer than nearly all of the 3FGL power law sources. At low energies where we find the most substantial support for the source, contamination from the diffuse or isotropic backgrounds is expected to increase. This soft spectrum may be suggestive of background contamination, rather than reflecting emission from the source star itself. 

\par In order to better understand the extraction of sources with soft energy spectra and the impact of background contamination, we use the wavelet analysis algorithm {\it PGWave}~\citep{1997ApJ...483..350D} to model the \eri{} sky region. Here we provide a brief description of {\it PGWave} and refer to the original literature for further details.

\begin{figure}
\centering
\includegraphics[width=\linewidth]{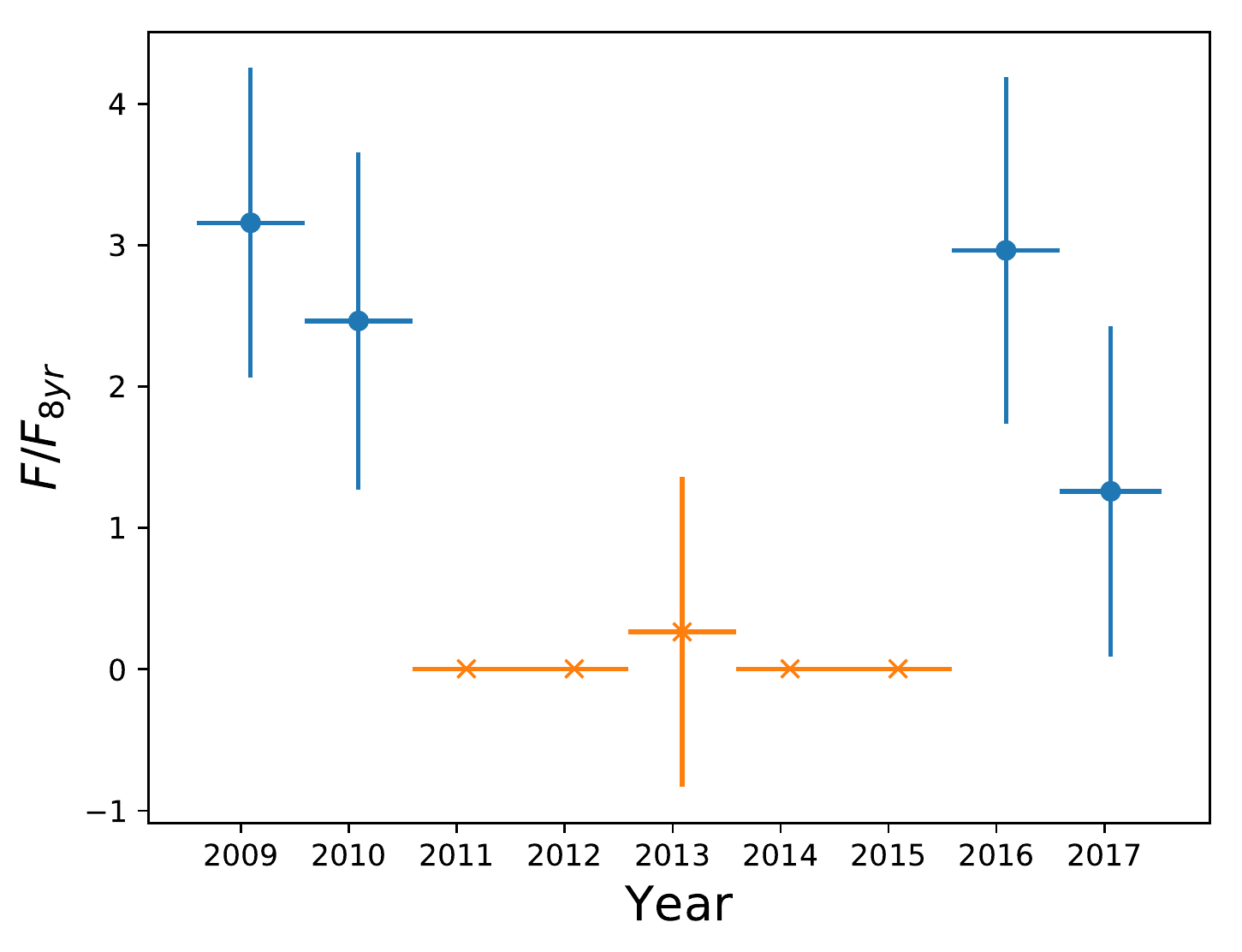}	
\caption{Test for variability of $\gamma$-ray emission from \eri{}. The fluxes shown are from a \texttt{gtlike} analysis after breaking the data into nine approximately one-year bins, normalized to the 8.9 year result. Orange crosses correspond to bins where $TS \lesssim 0$, blue circles are points where this was not the case.}
\label{fig:variability}
\end{figure}

\par Given a map of photon counts over a given energy range in a region of sky, the algorithm identifies overdensities in photon counts, or seeds, which are point-source candidates. The threshold for identifying these seeds is subject to an input signal-to-noise criterion based on the assumption of a locally constant background, without the assumption of an interstellar emission model. For our analysis we use a signal-to-noise criterion of 3$\sigma$. The seeds that are identified via the above procedure may be true point sources, or they may be structures in the interstellar emission that are indistinguishable from point sources due to the finite angular resolution and statistics of the Fermi-LAT data.

\begin{figure*}
\subfigure{\includegraphics[width=.49\linewidth]{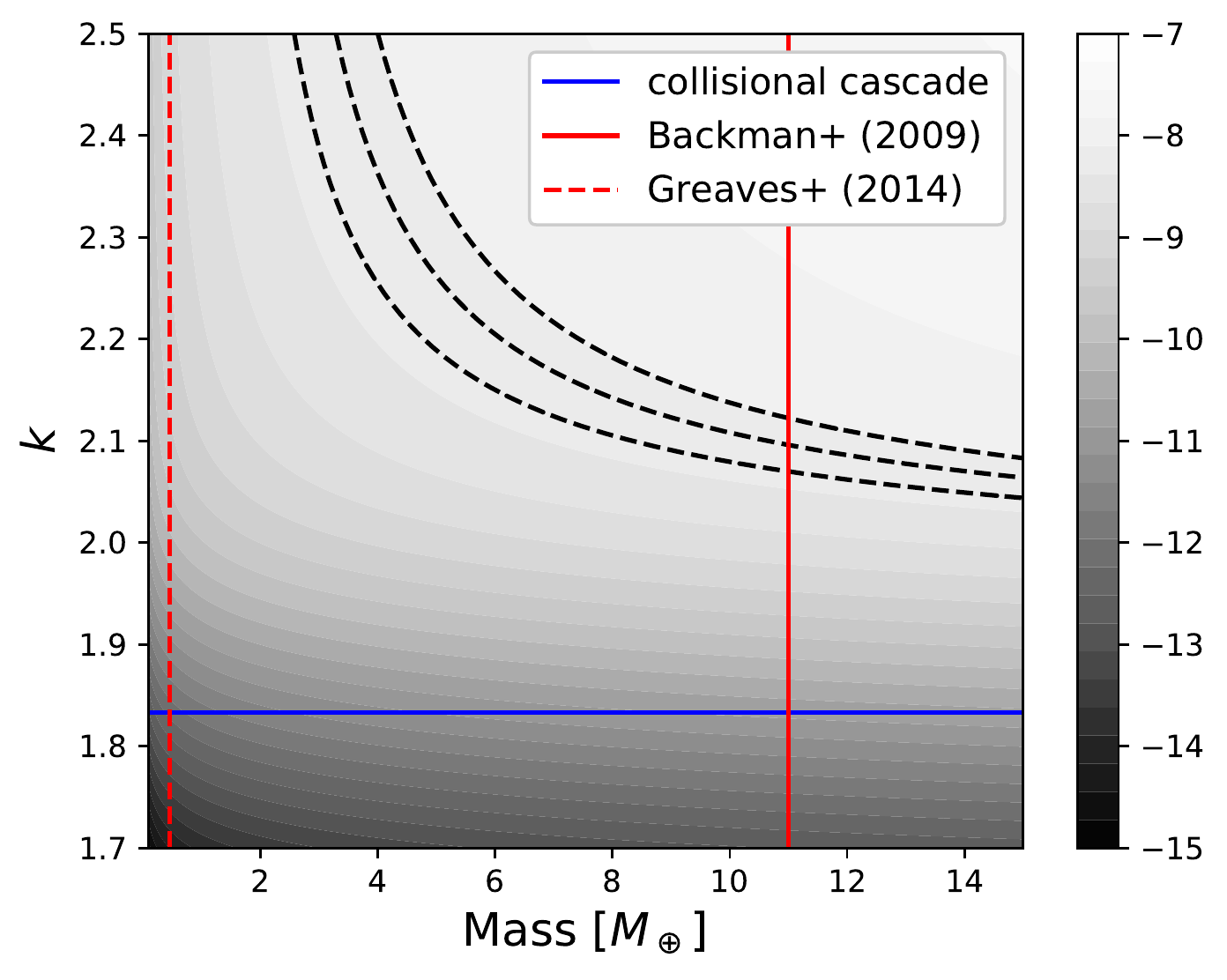}}
\subfigure{\includegraphics[width=.49\linewidth]{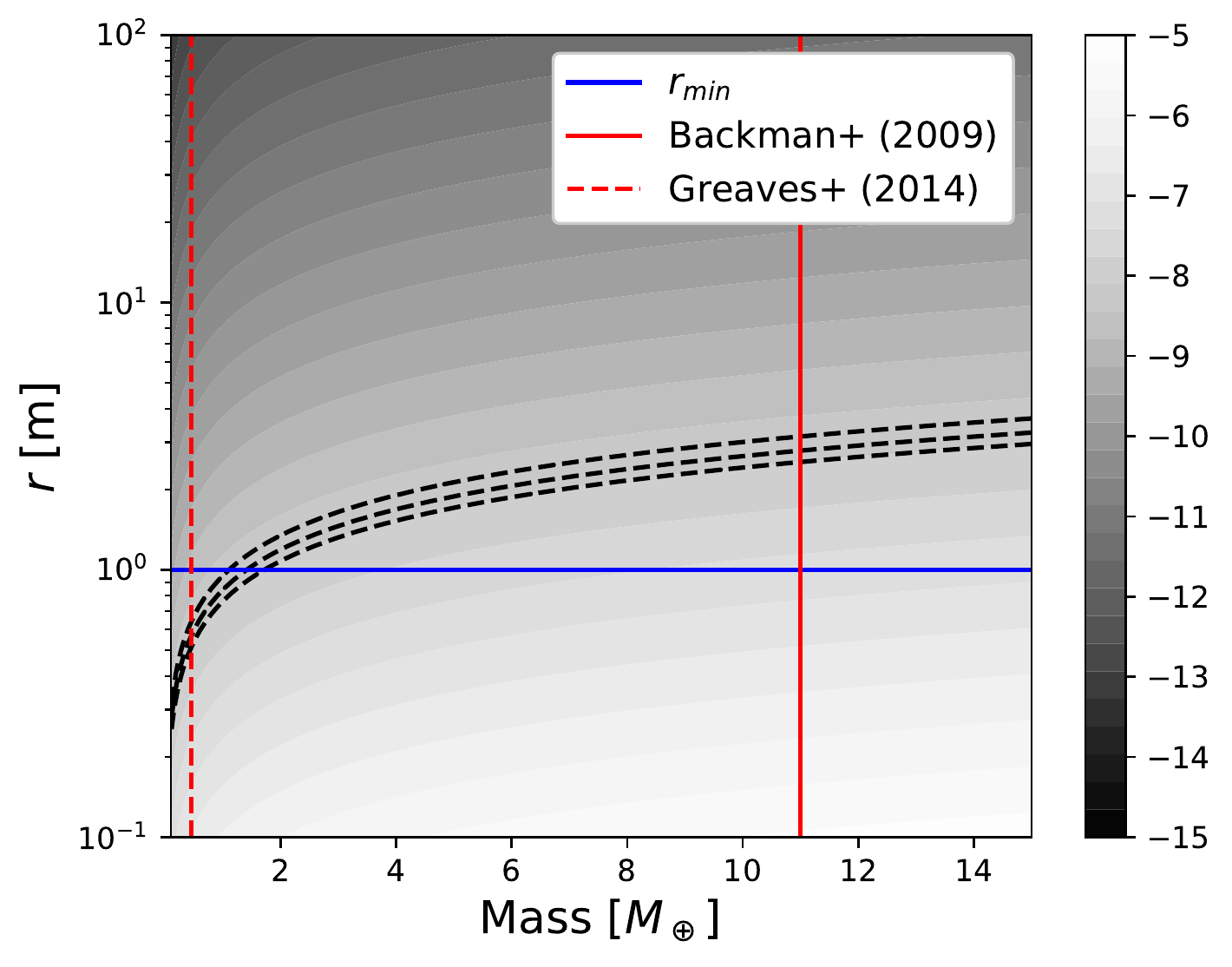}}
\caption{Debris disk interpretation of the \eri{} gamma ray signal. Contours show $\log_{10}(F)$, where the flux is in units of cm$^{-2}$ s$^{-1}$. The central black dashed line corresponds to the flux measurement, and the outer black dashed lines corresponds to the 1-$\sigma$ errors. Red lines show indirect mass measurements for the disk from \cite{2014ApJ...791L..11G} and \cite{2009ApJ...690.1522B}. Left: cascading model, blue line is commonly assumed collisional cascade $k=11/6$. Right: simple model made of bodies with uniform density and radius $r$, blue line is 1 meter (penetration depth for which the Moon provides a good template).
\label{fig:ddflux}
}
\end{figure*}

\par A significant complication that impacts the identification and localization of point sources is the relatively poor Fermi-LAT PSF at energies $\sim$100\,MeV. To account for the larger PSF at low energies, we employ the following algorithm in our identification of sources within a 10$^\circ$ region around \eri{}. We first break up the photons into three energy bins: 100-316 MeV, 316-1000 MeV, and 1-3.16 GeV. We then run {\it PGWave} on the ROI considering only photons in the highest energy bin, 1-3.16 GeV. The benefit of starting at high energy is that in this regime the 68\% containment of the PSF is $\sim 0.9^\circ$, so the seeds are well localized. With the locations of the high energy seeds identified, we run \texttt{gtlike} on this ROI to obtain the best fitting spectral energy distribution of the sources including the interstellar emission model. 

\par Using the locations and spectral energy distributions of the seeds identified at high energies and the interstellar emission model, we then run {\it PGWave} on the residuals for the same ROI in the two lower energy bins, $100-316$ MeV and $316-1000$ MeV. We merge the output seeds from these two low energy bins by identifying seeds that are within 1 degree of each other. When such overlapping seeds are identified, we use the position of the seed from the higher energy bin. From the combined seeds from these two energy bins and the interstellar emission model, we again run \texttt{gtlike} to obtain the significance of the seeds. The end result is an identification of seeds with significant $TS$ values that are candidate point sources in the region. 

\par The results of this analysis are shown in Figure \ref{fig:pgwave}. We find a seed at the location of \eri{} with a TS $\sim 25$, which is consistent with what was obtained from the \texttt{gtlike} analysis above. We are also able to recover 3 of the 4 3FGL sources within the region, which are indicated as white circles in Figure~\ref{fig:pgwave}. 

\subsubsection{Variability}

\par We now move on to test for variability of the source over the course of the 8.9 year data sample.\footnote{Note that the gamma-ray flux from the Sun appears to be variable \citep{Ng:2015gya,2018PhRvL.121m1103L}.} We break the data up into nine approximate one year time intervals and perform a \texttt{gtlike} analysis on each of the nine bins. This binning is motivated by the observed variability timescale for the magnetic field~\citep{2015AN....336..258L}. We begin by fixing the power law index at the best-fit value obtained above from the steady-state analysis, as well as fixing the diffuse and isotropic normalizations. This is justified by noting that the diffuse and isotropic emission should be constant across all of the nine bins, and the spectral shape from the steady-state analysis is characteristic of a cosmic-ray induced process. The only parameter allowed to vary is the normalization of the source at the location of \eri{}. 

\par For the above model, the fluxes are shown in Figure \ref{fig:variability}. There is weak indication of a high-low-high flux state, with a characteristic variability timescale of $\sim 7$ years. Defining $L_\imath^{steady}$ as the likelihood calculated in the $\imath^{th}$ time bin using the best-fitting normalization from the 8.9 year steady-state analysis, and $L_\imath^{var}$ as the likelihood calculated in the $\imath^{th}$ time bin allowing for a free normalization of the source, we find 
\begin{equation}
\sum_{\imath=1}^9 \left(2 \log L_\imath^{var} - 2 
\log L_\imath^{steady} \right) = 14.9. 
\end{equation}
This corresponds to a p-value = 0.05, or $\lesssim 2\sigma$ indication of source variability. We have additionally confirmed a similar flux pattern for cases in which the diffuse normalizations and the spectral index for the \eri{} sources are allowed to vary in each of the time bins, with small changes to the measured flux values.

\subsubsection{Inverse Compton}
\par We perform an additional test to verify that the gamma-ray emission from \eri{} is best described by a soft spectrum, as opposed to hadronic emission from the stellar disk or from IC scattering of cosmic-ray electrons from stellar photons. Using the code developed by~\citet{2006ApJ...652L..65M}, we estimate the gamma-ray spectrum from IC scattering, assuming a luminosity of $L = 0.34 L_\odot$ and $T_{eff} = 5084$ K for \eri{}, and  a conservative interstellar spectrum which gives the maximum IC emission. With the IC spectrum, for all background sources fixed at their 3FGL values, we find that the TS for \eri{} drops to $\sim$7, as compared to the TS values is Table~\ref{table:checks}. We are thus able to conclude that the emission from \eri{} is better described by a soft power law spectrum than a harder spectrum predicted by IC emission. 

\subsubsection{Nearby point sources} 

\par As discussed in \citet{2015JCAP...09..016H}, a significant fraction of high $TS$ points in the ``blank sky'' correspond to unresolved blazars, radio galaxies, and star forming galaxies. This motivates us to search for sources near \eri{} that are detected at other wavelengths, but do not have any associated gamma-ray emission. Here we consider sources that are within 1.5$^\circ$ of \eri{} and examine several catalogs at other wavelengths to determine if any sources in these catalogs overlap with \eri{}: the Roma-BZCAT Multi-Frequency Catalog of Blazars (BZCAT) \citep{2015Ap&SS.357...75M}, the Combined Radio All-Sky Targeted Eight-GHz Survey (CRATES) catalog \citep{2007ApJS..171...61H}, the Candidate Gamma-Ray Blazar Survey (CGRaBS) catalog \citep{2008ApJS..175...97H}, and the Australia Telescope National Facility (ATNF) pulsar catalog \citep{2005AJ....129.1993M}.

\par As a result of our search we find two radio-bright blazars in the CRATES catalog  located within 1.5$^\circ$ of \eri{}: CRATES J033149-105155 and CRATES J032952-100251. A re-analysis of the binned likelihood including point sources at the locations of both blazars with a fixed power law index of $\Gamma = 2.2$, near both the median and mean value for a blazar in the 3FGL catalog, yields $TS < 2.5$ for both blazars and an \eri{} $TS \sim 21$.  This, combined with the fact that the best fit power law $\Gamma \simeq 3.6$ is well outside of the normal range of spectral indices for blazars, suggests that the signal is unlikely to originate from overlapping blazars in the CRATES catalog.

\section{Interpretation and Discussion}
\label{sec:discussion}

\par Under the assumption that the gamma-ray emission is associated with~\eri{}, in this section we present two plausible theoretical interpretations for this emission: (1) from the solid bodies in the debris disk and (2) emission from the stellar activity. 

\subsection{Debris disk model}

\begin{figure*}
\centering
\subfigure[\ceti{}]{\label{fig:ceti_cascade}
\includegraphics[width=.32\linewidth]{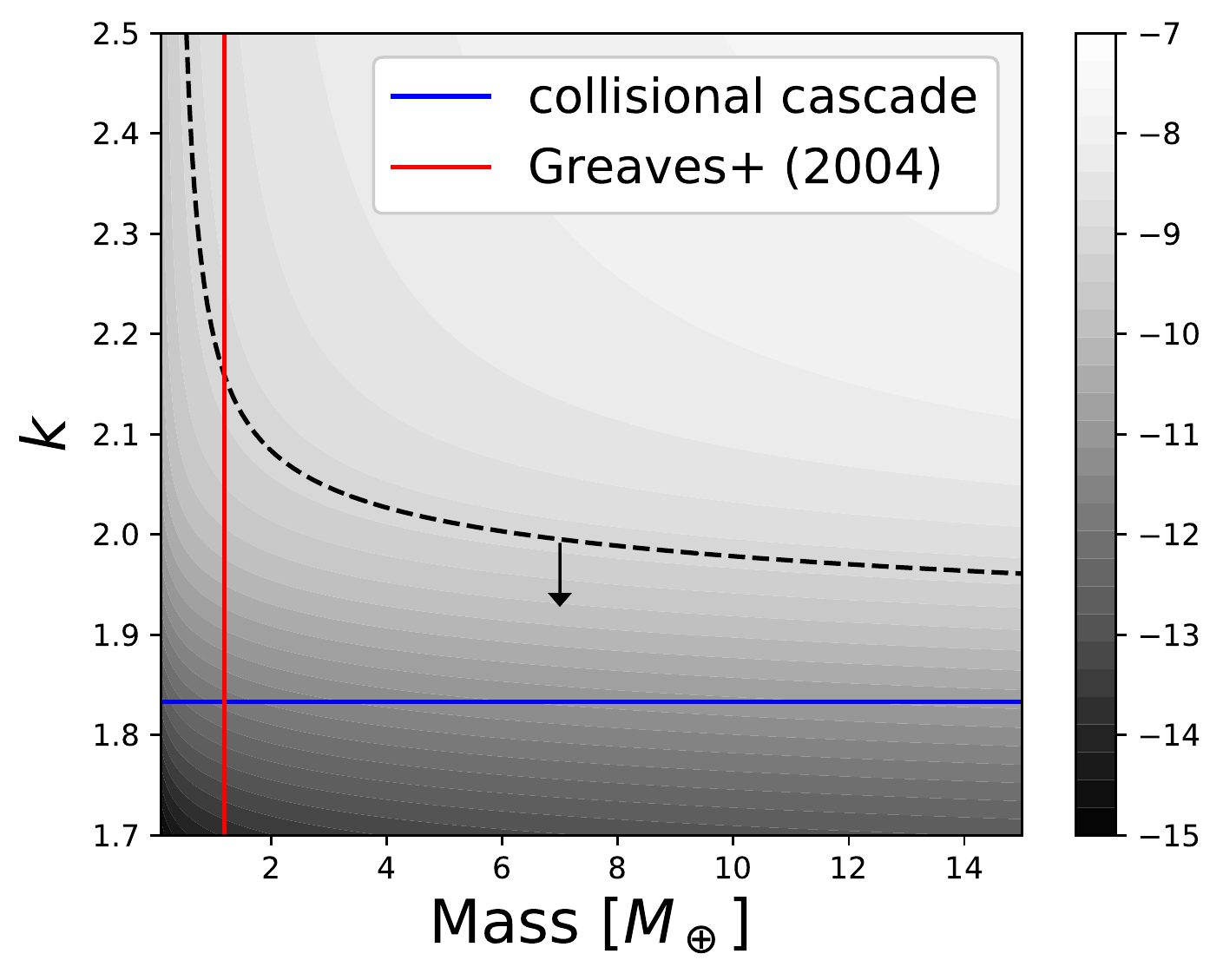}}
\subfigure[Fomalhaut]{\label{fig:fomalhaut_cascade}
\includegraphics[width=.32\linewidth]{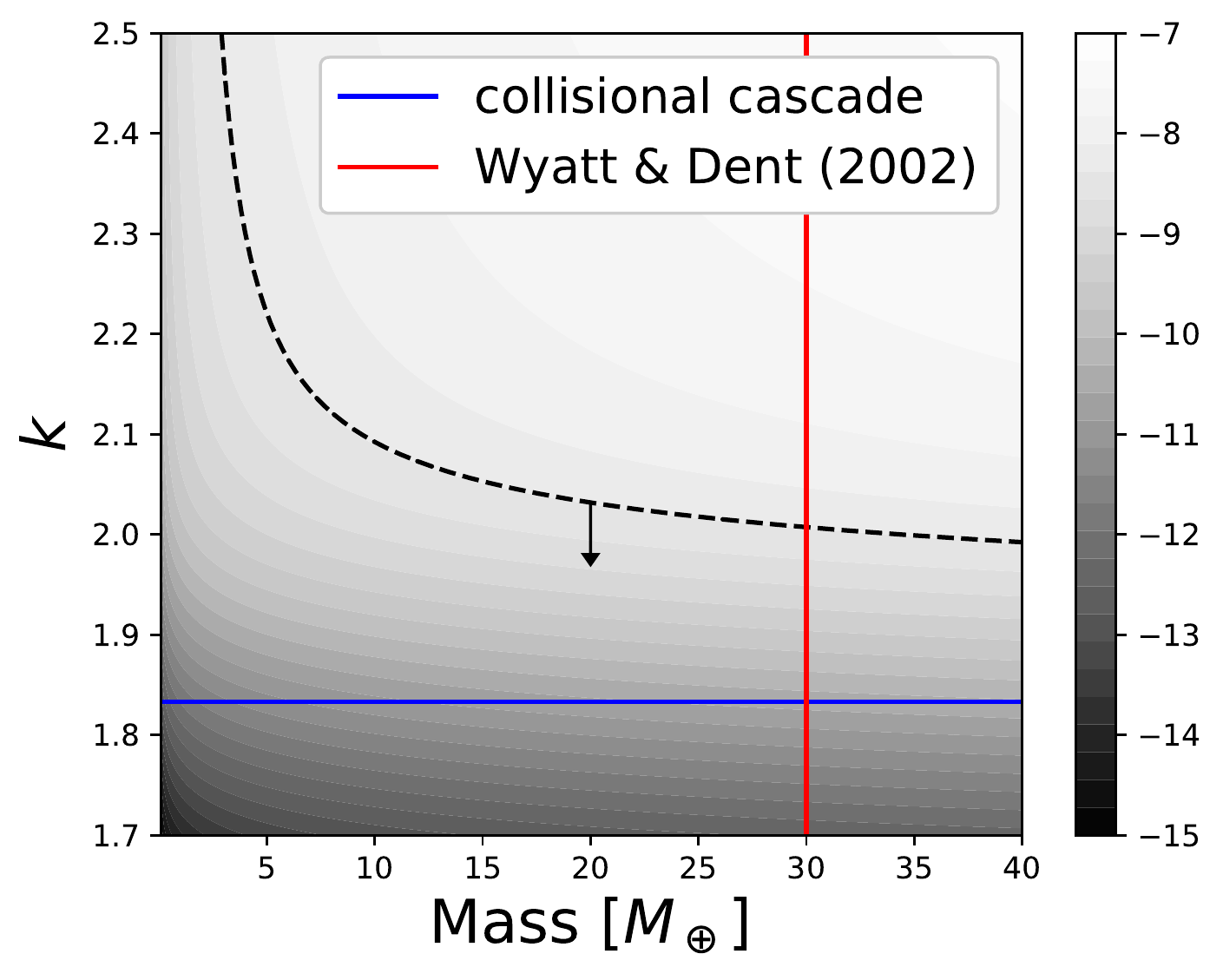}}
\subfigure[Vega]{\label{fig:vega_cascade}
\includegraphics[width=.32\linewidth]{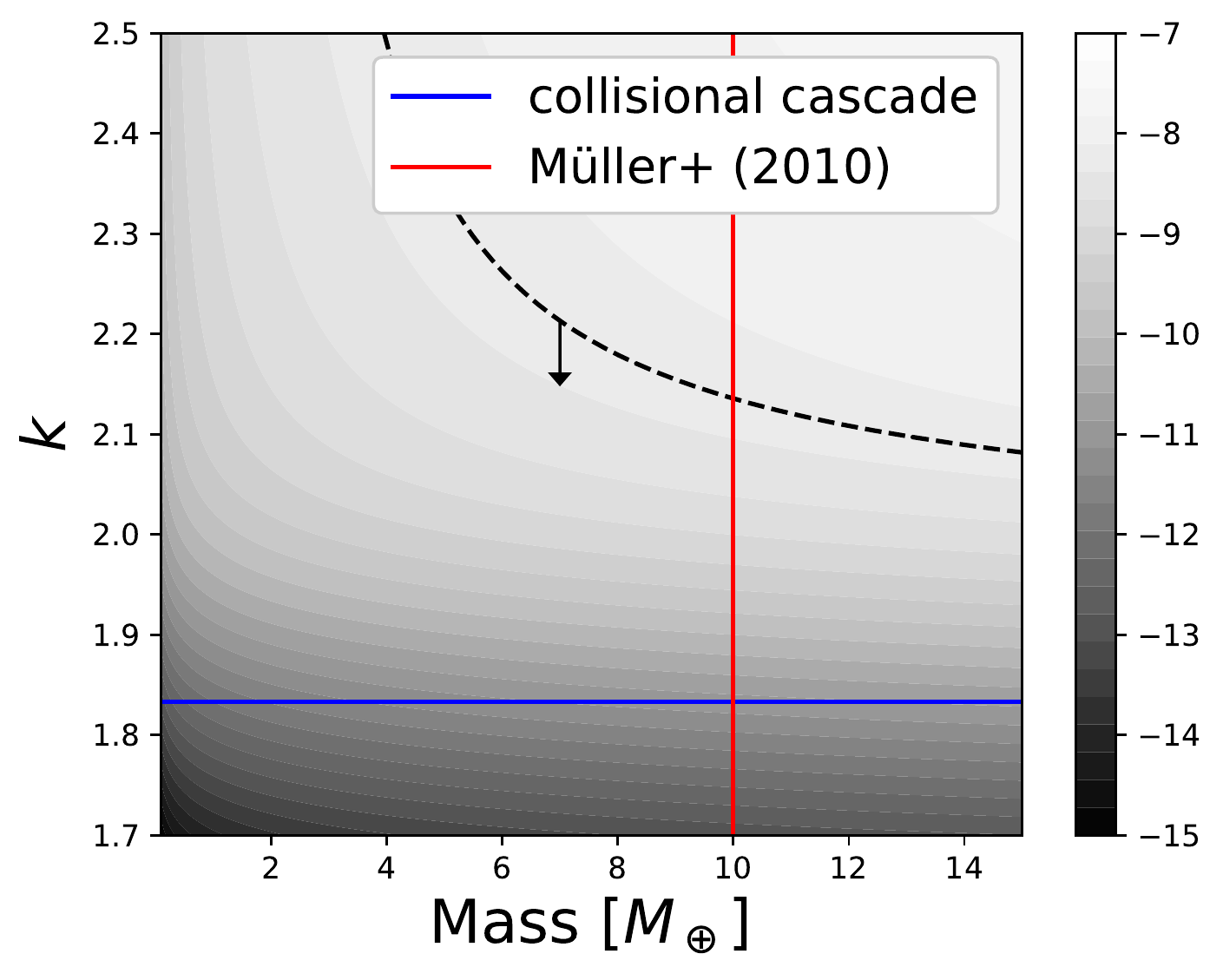}}
\caption{Upper limits on the size distribution power law index $k$ as a function of the total mass in the debris disk for $\tau$ Ceti, Fomalhaut, and Vega. Red lines show indirect mass measurements for the disk from the references indicated. Blue lines indicate the collisional cascade model. Regions above the dashed black curves are excluded by the gamma-ray data.}
\label{fig:limits}
\end{figure*}

The soft gamma-ray spectrum may indicate that the origin of the gamma-ray emission is from solid bodies in the debris disk. Including the bias towards softer sources as described above in the \texttt{gtobssim} simulations, our best-fit spectral index is harder than the $> 200$ MeV lunar spectrum.  Assuming that the emission originates from the disk, we can constrain the size distribution of bodies that comprise this disk and compare to the corresponding distribution obtained from the IR emission. Figure \ref{fig:ddflux} shows contours of the gamma-ray flux  in the ($k-M$) plane, where $k$ is the power-law index defined in Section~\ref{sec:debrisdisks} and $M$ is the total mass of the debris disk. Figure~\ref{fig:ddflux} shows that interpreting the flux as from the debris disk, the value of $k$ is greater than that expected for the collisional cascade model, which is shown as the blue horizontal line in the left panel of Figure~\ref{fig:ddflux}. The gamma-ray flux is consistent with the IR measurements from~\citet{2009ApJ...690.1522B} for an inferred total debris disk mass of $\sim 11 M_\oplus$ \citep[though see][who infer $\sim 0.45 M_\oplus$]{2014ApJ...791L..11G}.

\par In the right panel of Figure~\ref{fig:ddflux}, we assume the simple model that all of the bodies in the debris disk are the same size. In this case, the gamma-ray flux measurement is consistent with the respective IR mass measurements for sizes $\sim 0.5-0.6$m \citep[using][]{2009ApJ...690.1522B} and $\sim 2-3 $m \citep[using][]{2014ApJ...791L..11G}.

\par From the null gamma-ray detections from $\tau$ Ceti, Fomalhaut, and Vega, limits may be obtained on the properties of their debris disk. Bounds on the combination of the power law index $k$ and the total mass in the disk are shown in Figure~\ref{fig:limits}. For all three stars, the gamma-ray data are able to exclude a range of $k \gtrsim 2$ that are steeper than the collisional cascade model. 

\par We note that the limits in Fig.~\ref{fig:limits} and uncertainty bands shown in Fig.~\ref{fig:ddflux} assume the lunar gamma-ray spectrum, and a size cut-off of $1$m for the bodies in the disk. Extrapolating to lower size dust grains would require modeling the spectral energy distribution from this component, which would introduce further uncertainties in the bands in Fig.~\ref{fig:ddflux} and the bounds in Fig.~\ref{fig:limits}. 

\subsection{Stellar Activity}

\par A variable gamma-ray source securely identified with \eri{} would challenge our debris disk interpretation. In this case there could be two interpretations. The first is that the stellar wind, which is known to be variable, modulates the cosmic ray intensity within its termination shock which lies roughly $\sim1000$ AU from the star, well beyond the observed disk. We know that the solar wind, which has a momentum flux $\sim0.03$ that of the \eri{} stellar wind, suppresses the Galactic cosmic ray flux at Earth by roughly an order of magnitude at energies $\sim$ a few hundred MeV \citep[e.g.][]{2017PhRvD..96j3005T}. An even larger suppression should be expected in \eri{}. The stellar wind can just cover the observed disk in the putative $\sim3$ yr variation timescale but much faster variation could not be interpreted in this fashion.

\par The second possible interpretation is that the gamma rays originate in the stellar corona, not the disk. \eri{} has an age, spin period, and average surface magnetic field $\sim$ 0.1, 0.4, 10 times these quantities for the Sun so we certainly expect its corona to be much more active than the solar corona. However, from the measured flux the gamma-ray luminosity of \eri{} is derived to be $\sim 10^{27}$ erg s$^{-1}$, corresponding to $\sim10^{10}$ that of the quiet sun and $\sim10^7$ that of a powerful solar flare. \Edit{\eri{} has also been studied in X-rays, which indicate strong coronal activity \citep[\eri{}'s x-ray luminosity is roughly 10 times that of solar;][]{1981ApJ...243..234J}, and in radio, which indicates steady-state emission $\ge 8$ MHz that is consistent with a stellar origin~\citep{2018ApJ...857..133B}.} If there were such an  enormous increase in coronal activity, then we would also expect to see  powerful radio emission and dramatic, rapid variability and we do not do so. We find that coronal emission is unlikely to explain the gamma-ray flux coincident with \eri{}.

\section{Conclusions} 
\label{sec:conclusions}
\par We have used the Fermi-LAT to search for gamma-ray emission around four nearby main sequence stars with debris disks. We find tentative evidence for a gamma-ray signal around \eri{} with a $TS \sim 25$, with the precise value depending on the energy cuts and background modeling that is assumed. However, because the results from the photon PSF cuts may point to an extended emission in this region, we are unable to rule out that this emission is due to a more extended feature in the diffuse background. If ultimately confirmed as being due to~\eri{} itself, this would be the first indication of gamma-ray emission from the vicinity of a main sequence star other than the Sun.  

\par If the emission originates from the debris disk, our analysis could provide a new avenue for studying debris disks around nearby stars. It represents the first proposed method to study debris disks in a wavelength band other than the IR, which is only sensitive to the dust component in a disk. Furthermore, gamma rays provide the only proposed means to more directly study the size distribution of solid bodies in a debris disk. The flux from \eri{} is consistent with the emission from the debris disk if the size distribution is dominated by $\sim$1m-sized bodies. The detected flux is greater than the predictions of the collisional cascade model. For the remaining three stars that we study (\ceti{}, Fomalhaut, Vega), the upper limits on the gamma-ray flux are consistent with the predictions of the collisional cascade model.

\par The analysis that we have proposed can be improved upon from several perspectives, both theoretically and observationally. On the theoretical side, to model the IR observations we have used the simple black-body model for the IR emission and a simple power law model for the size distribution of bodies in the disk. Also, due to non-uniform winds from the stars and possibly the presence of giant planets, our assumption that the cosmic ray flux that is similar around all disks is likely an oversimplification. \Edit{On the observational side, as more data is collected, it is possible that the tentative signal from~\eri~will increase (or decrease) in significance. Though we anticipate the change is not likely to be too substantial, since for point sources near the detection threshold Fermi-LAT is already in the systematics-dominated regime due to diffuse and isotropic backgrounds. Also as an extension, one could consider a stacking analysis of the gamma-ray data around all nearby stars that harbor known debris disks.} Debris disks will also contribute to the diffuse IR and gamma-ray backgrounds; an reliable estimate of these fluxes would require a more detailed spectral model of the gamma rays from the disks. 
 
\par Shedding more light on debris disks is also important for elucidating how they interplay with planetary formation. In the particular case of \eri{}, there are indications of an approximate Jupiter-sized planet from radial velocity measurements, though firm establishment of a planet appears inconclusive~\citep{2012ApJS..200...15A}. We finally note that this high a flux of gamma rays, which cannot be shielded by a magnetosphere, would not be conducive to the development of life, as was once considered to be a possibility for \eri{}.\footnote{https://www.seti.org/seti-institute/project/details/early-seti-project-ozma-arecibo-message}
 
\section*{Acknowledgements}

\par We thank Andy Strong, Regina Caputo, \Edit{and an anonymous referee} for a detailed reading of this paper and for several valuable suggestions. This work was supported by NSF grant AST-1263034, ``REU Site: Astronomical Research and Instrumentation at Texas A\&M University.'' We acknowledge support from NASA Astrophysics Theory grant NNX12AC71G. Work at NRL is supported by NASA.

\par The \textit{Fermi} LAT Collaboration acknowledges generous ongoing support
from a number of agencies and institutes that have supported both the
development and the operation of the LAT as well as scientific data analysis.
These include the National Aeronautics and Space Administration and the
Department of Energy in the United States, the Commissariat \`a l'Energie Atomique
and the Centre National de la Recherche Scientifique / Institut National de Physique
Nucl\'eaire et de Physique des Particules in France, the Agenzia Spaziale Italiana
and the Istituto Nazionale di Fisica Nucleare in Italy, the Ministry of Education,
Culture, Sports, Science and Technology (MEXT), High Energy Accelerator Research
Organization (KEK) and Japan Aerospace Exploration Agency (JAXA) in Japan, and
the K.~A.~Wallenberg Foundation, the Swedish Research Council and the
Swedish National Space Board in Sweden.
 
\par Additional support for science analysis during the operations phase is gratefully acknowledged from the Istituto Nazionale di Astrofisica in Italy and the Centre National d'\'Etudes Spatiales in France. This work performed in part under DOE Contract DE-AC02-76SF00515.

\Edit{\software{Astropy \citep{astropy13, astropy18}, Jupyter (\href{https://jupyter.org/}{jupyter.org}), Matplotlib \citep{matplotlib}, NumPy \citep{numpy}, Pandas \citep{pandas}, SciPy \citep{scipy}.}}

\appendix
\par In Section~\ref{sec:results}, we have allowed the 3FGL sources within 5 degrees of our sample of stars to vary in the fits. In Tables~\ref{table:5degallstars} and~\ref{table:5degchecks} we compare the fluxes that we obtain for the 3FGL sources and compare our fluxes to those from the 3FGL catalog. Table~\ref{table:5degallstars} reports the fluxes from the analysis in Table~\ref{table:allstars}, and Table~\ref{table:5degchecks} reports the fluxes from the analysis in Table~\ref{table:checks}.  

\begin{table}[h!]
\begin{center}
\caption{Comparison between 3FGL and 5DEG fluxes (in $10^{-9}$ ph cm$^{-2}$ s$^{-1}$) for each 3FGL source within 5$^\circ$ of each star, using the analyses from Table \ref{table:allstars}. 3FGL fluxes are computed by adding the flux values in five energy bands (with edges 100 MeV, 300 MeV, 1 GeV, 3 GeV, 10 GeV, 100 GeV). 3FGL flux errors are computed by adding the errors in each of the same bands; in the case where the lower flux error in a single band is greater than the flux value in that band (ie. a negative value for lower limit on the flux in that band), the flux value in that band is used instead.}
\begin{tabular}{ccccccccc}
\tableline\tableline
Star & 3FGL Name & 3FGL Flux & 5DEG Flux \\
\tableline
\multirow{4}{*}{\eri{}} & J0315.5-1026 & $10.0^{+4.0}_{-3.9}$ & $16.3 \pm 1.1$ \\
& J0324.5-1315 & $8.7^{+3.7}_{-3.6}$ & $5.5 \pm 1.1$ \\
& J0336.9-1304 & $13.6^{+3.9}_{-3.7}$ & $8.1 \pm 1.2$ \\
& J0349.2-1158 & $1.0^{+2.6}_{-0.8}$ & $1.4 \pm 0.2$ \\
\tableline
\ceti{} & J0132.6-1655 & $48.1^{+3.4}_{-3.4}$ & $39.3 \pm 1.1$ \\
\tableline
\multirow{3}{*}{Fomalhaut} & J2248.6-3235 & $13.3^{+3.2}_{-3.1}$ & $15.2 \pm 1.2$ \\
& J2250.7-2806 & $57.5^{+4.9}_{-4.9}$ & $44.9 \pm 1.1$ \\
& J2258.0-2759 & $92.9^{+4.7}_{-4.7}$ & $65.6 \pm 1.3$ \\
\tableline
\multirow{3}{*}{Vega} & J1820.3+3625 & $4.1^{+2.9}_{-2.9}$ & $1.4 \pm 0.2$ \\
& J1824.4+4310 & $0.3^{+3.6}_{-0.2}$ & $0.8 \pm 0.2$ \\
& J1838.1+3827 & $15.2^{+3.5}_{-3.4}$ & $10.7 \pm 4.0$ \\
& J1848.9+4247 & $6.2^{+2.8}_{-2.5}$ & $0.8 \pm 0.2$ \\
\tableline
\end{tabular}
\end{center}
\label{table:5degallstars}
\end{table}

\begin{table}[h!]
\begin{center}
\caption{Comparison between 3FGL and 5DEG fluxes (in $10^{-9}$ ph cm$^{-2}$ s$^{-1}$) for each 3FGL source near \eri{}, using the analyses from Table \ref{table:checks}. See Table \ref{table:5degallstars} for details on 3FGL flux errors.}
\begin{tabular}{ccccccccc}
\tableline\tableline
& J0315.5-1026 & J0324.5-1315 & J0336.9-1304 & J0349.2-1158 \\
3FGL & $10.0^{+4.0}_{-3.9}$ & $8.7^{+3.7}_{-3.6}$ & $13.6^{+3.9}_{-3.7}$ & $1.0^{+2.6}_{-0.8}$ \\
$>100$ MeV (fiducial) & $16.3 \pm 1.1$ & $5.5 \pm 1.1$ & $8.1 \pm 1.2$ & $1.4 \pm 0.2$ \\
\tableline
$>200$ MeV & $16.4 \pm 1.2$ & $6.3 \pm 1.3$ & $8.6 \pm 1.3$ & $1.5 \pm 0.2$ \\
$>500$ MeV & $16.1 \pm 1.4$ & $6.5 \pm 1.6$ & $8.5 \pm 1.8$ & $1.5 \pm 0.2$ \\
$>1$ GeV & $14.7 \pm 1.8$ & $7.1 \pm 2.1$ & $6.2 \pm 2.1$ & $1.5 \pm 0.2$ \\
\texttt{PSF0+PSF1} & $16.5 \pm 1.1$ & $6.1 \pm 1.1$ & $9.0 \pm 1.2$ & $1.3 \pm 0.2$ \\
\texttt{PSF2+PSF3} & $15.8 \pm 1.1$ & $5.5 \pm 1.1$ & $8.9 \pm 1.2$ & $1.5 \pm 0.2$ \\
\tableline
\end{tabular}
\end{center}
\label{table:5degchecks}
\end{table}

\bibliographystyle{aasjournal}	
\bibliography{refs,software} 	


\end{document}